\documentclass[twocolumn,prb,superscriptaddress]{revtex4-1}  
\pdfoutput=1

\usepackage{graphicx,bm,amsmath,amssymb,color}

\usepackage{hyperref}
\usepackage{cleveref}

\usepackage{amssymb}
\usepackage{lipsum}
\usepackage{braket}

\allowdisplaybreaks

\let\hide\iffalse

\def\up{{\uparrow}}
\def\down{{\downarrow}}

\def\bk{{\bf k}}

\def\br{{\bf r}}
\def\brp{{\bf r}^\prime}

\def\bq{{\bf q}}
\def\a{{\alpha}}

\def\bR{{\bf R}}

\def\btau{{\bm \tau}}
\def\D{\partial}
\def\d{\delta}
\def\w{\omega}
\def\bE{{\bf E}}

\def\bG{{\bf G}}
\def\bT{{\bf T}}
\def\bD{{\bf D}}

\def\<{\langle}
\def\>{\rangle}
\def\k{\kappa}
\def\ve{\varepsilon}
\def\e{\epsilon}

\def\wf{{\rm w}}
\def\ha{{\hat{a}}}

\def\hH{{\hat{H}}}

\begin{document}

\title{\textit{Ab initio} theory of polarons: formalism and applications}

\author{Weng Hong Sio} 
\affiliation{Department of Chemistry, Physical and Theoretical Chemistry,
University of Oxford, South Parks Road, Oxford, OX1 3QZ, UK}
\affiliation{Department of Materials, University of Oxford, Parks Road, Oxford, OX1 3PH, UK} 
\author{Carla Verdi} 
\altaffiliation[Present address: University of Vienna, Faculty of Physics and Center for
Computational Materials Sciences, Sensengasse 8/12, 1090 Vienna, Austria]{}
\author{Samuel Ponc\'e} 
\author{Feliciano Giustino}
\email{feliciano.giustino@materials.ox.ac.uk}
\affiliation{Department of Materials, University of Oxford, Parks Road, Oxford, OX1 3PH, UK}
\date{\today}

\begin{abstract} 
We develop a theoretical and computational framework to study polarons in semiconductors and insulators
from first principles. Our approach provides the formation energy, excitation energy, and wavefunction of both electron and hole polarons, and takes into account the coupling of
the electron or hole to all phonons. An important feature of the present method is that it does not
require supercell calculations, and relies exclusively on electron band structures, phonon
dispersions, and electron-phonon matrix elements obtained from calculations in the crystal unit cell.
Starting from the Kohn-Sham (KS) equations of density-functional theory, we formulate the polaron problem
as a variational minimization, and we obtain a nonlinear eigenvalue problem in the basis of KS 
states and phonon eigenmodes. In our formalism the electronic component of the polaron is expressed
as a coherent superposition of KS states, in close analogy with the solution of the Bethe-Salpeter
equation for the calculation of excitons. We demonstrate the power of the methodology by studying polarons
in LiF and Li$_2$O$_2$. We show that our method describes both small and large polarons, and seamlessly 
captures Fr\"ohlich-type polar electron-phonon coupling and non-Fr\"ohlich coupling to acoustic and optical
phonons. To analyze in quantitative terms the electron-phonon coupling mechanisms leading to the formation 
of polarons, we introduce spectral decompositions similar to the Eliashberg spectral function.
We validate our theory using both analytical results and direct calculations on large 
supercells. This study constitutes a first step toward complete {\it ab initio} many-body calculations 
of polarons in real materials.
\end{abstract}

\maketitle

\section{Introduction}\label{sec.intro}

The polaron is a quasiparticle that can be found in many crystalline solids such as 
semiconductors,\cite{Lindemann1983} insulators,\cite{Popp1972} and molecular crystals.\cite{Chaikin1972} 
A polaron is formed when an electron or a hole couples to the ions in a crystal in such a way 
as to generate a lattice distortion; the distortion in turn produces an electric field that acts on the electron or 
hole. This feedback mechanism alters the energetics and dynamics of the charge carrier and may induce self-trapping. With the
improvement in the energy and momentum resolution of angle-resolved photoelectron spectroscopy (ARPES),
it has become possible to probe these quasiparticles in many systems of interest, from
transition metal oxides,\citep{Moser2013,Cancellieri2016,Riley2018,Wang2016} to two-dimensional 
materials.\citep{Chen2015,Asensio2018,Kang2018} These experiments and related theoretical investigations 
contributed to reinvigorating the interest in polaron physics.\citep{Verdi2017,Riley2018,Gonze2018,
Antonius2015}

The notion of polaron was introduced in a classic short paper by Landau,\cite{Landau1933} and quantitative 
studies started with the work of Pekar,\cite{Pekar1946} who considered a single electron interacting
with a dielectric continuum. This interaction was shown to induce a localization of the wavefunction, 
and an enhancement of the effective mass of the electron.\cite{Landau1948} Shortly afterwards, 
Fr\"ohlich, Pelzer, and Zienau formulated a quantum-mechanical theory of the polaron, where the 
interaction with the polarizable continuum was replaced by electron-phonon interactions (EPIs) 
between the excess electron and the longitudinal optical phonons of the lattice.\cite{Frohlich1950} 
Subsequent work by Lee, Low, and Pines,\cite{Lee1953} Fr\"ohlich,\cite{Frohlich1954} 
Feynman,\cite{Feynman1955} and others,\cite{Gross1955,Marshall1970,Luttinger1980,Kholodenko1983,
Bogolubov2014} focused on determining accurate solutions of the Fr\"ohlich polaron Hamiltonian
for various strengths of the EPI. More recent work includes accurate numerical investigations of the 
Fr\"ohlich Hamiltonian using the diagrammatic Monte Carlo method,\cite{Prokofev1998} path-integral 
Monte Carlo,\cite{Titantah2001} and the renormalization group approach.\cite{Grusdt2016} For 
comprehensive and up-to-date reviews of this vast research area we refer the reader to 
Refs.~\onlinecite{Alexandrov2010,Grusdt2015,Devreese2016}. 

Despite the successes of these model solutions and the growing interest in applying these techniques 
to novel areas such as ultracold atoms,\cite{Tempere2009,Rath2013,Grusdt2016} the
Fr\"ohlich Hamiltonian describes a highly-idealized model system, and does not contain enough information
to begin a quantitative and predictive study of polarons in real solids. In fact, this model
considers the coupling of an electron to a dispersionless longitudinal optical phonon, but in most
materials of practical interest the EPI is far more complex. For example halide perovskites such as
CH$_3$NH$_3$PbI$_3$ exhibit multi-phonon Fr\"ohlich coupling,\cite{Sendner2016,Schlipf2018,Ponce2019} and 
transition metal oxides such as TiO$_2$ exhibit anisotropic effective masses.\cite{Verdi2017} 
Furthermore in many situations the EPI involves both long-range and short-range effects, which are 
not well captured by the two limiting scenarios investigated by Fr\"ohlich\cite{Frohlich1950} and 
Holstein \cite{Holstein1959}. In order to mitigate these drawbacks, considerable effort is being 
devoted to expanding the scope of model Hamiltonians to additional EPI mechanisms.\cite{Vlietinck2015} 
In our view what is still missing in this area is a unified approach to the polaron problem, 
where the EPI mechanisms and parameters are obtained from first principles, without making 
\textit{a priori} assumptions.

An obvious candidate for beginning to develop an \textit{ab initio} theory of polarons is density-functional
theory (DFT). However, DFT studies of polarons also carry some limitations. Since the calculations are
performed by adding or removing an electron in a supercell, the computational cost restricts the systems
that can be investigated to small- and intermediate-size polarons (i.e. supercells containing up to a
few thousand atoms).\cite{Spreafico2014} This limitation makes it difficult to investigate systems
with interesting long-range Fr\"ohlich EPIs.\cite{Verdi2015} On top of these computational challenges, 
standard DFT calculations suffer from the self-interaction error,\cite{Zunger1981} and this can be 
critical in the study of polarons. Several promising attempts at circumventing this problem have been made, 
ranging from using Hubbard-corrected DFT\cite{Setvin2014,Himmetoglu2014}, to hybrid 
functionals,\cite{Setvin2014,Himmetoglu2014,kokott2018} and specialized self-interaction correction 
(SIC) schemes.\cite{Sadigh2015} Even though it is reasonable to expect that these technical challenges 
will be overcome in the future, DFT calculations based on supercell calculations offer limited physical 
insight into the EPI mechanisms that drive polaron formation. As a result, it is difficult to establish 
a link between such calculations and more advanced many-body solvers for model Hamiltonians.

The goal of the present study is to make \textit{ab initio} DFT calculations of polarons more accessible
and more systematic, and to lay the groundwork for linking these calculations with advanced polaron
solvers based on model Hamiltonians. To this aim we reformulate the calculation of polaron energies
and wavefunctions using DFT and supercells into a nonlinear eigenvalue problem. The ingredients of this
nonlinear problem are DFT quantities that are obtained exclusively from calculations in the crystal
unit cell, namely electron bands, phonon dispersions, and electron-phonon matrix elements; the method
does not require explicit supercell calculations. Our present approach is similar in spirit to the
study of excitons via the Bethe-Salpeter equations:\citep{Rohlfing2000,Bokdam2016} as in the exciton problem,
we write the polaron wavefunction as a superposition of Kohn-Sham (KS) states, and we seek to determine
the expansion coefficients in this basis. This is achieved by performing a variational minimization, 
and the resulting `polaron equations' are found to be closely related to the Landau-Pekar theory.
The key approximations involved in our approach are those of harmonic phonons and linear electron-phonon
coupling, as in the original Fr\"ohlich model and in the vast majority of modern many-body investigations
of polarons. We illustrate the capability of this new theoretical and computational framework by discussing
applications to the large electron polaron in LiF, the small hole polaron in the same compound, and the
small electron polaron in Li$_2$O$_2$. For these test cases we report polaron formation energies and 
excitation energies, wavefunctions, and atomic displacement profiles, and we analyze the underlying EPI
mechanisms in each case. We also discuss a self-interaction scheme that eliminates the need
for Hubbard corrections or hybrid functionals. A preliminary account of this work was given in 
Ref.~\onlinecite{Sio2018}.

The manuscript is organized as follows. In Sec.~\ref{sec.lp} we review the classic Landau-Pekar 
model.\cite{Landau1933,Pekar1946,Alexandrov2010} In Sec.~\ref{sec.abinitio} we develop our 
formalism. We start from the derivation of the polaron equations in Sec.~\ref{sec.pol-eqs}, 
we discuss the formation energy and the excitation energy in Sec.~\ref{sec.pol-en},
and we recast the problem in the basis of KS states and vibrational eigenmodes in Sec.~\ref{sec.pol-eqs-ks}.
In Sec.~\ref{sec.pol-eqs-h} we obtain the atomic displacement patterns associated with the polaron,
and in Sec.~\ref{sec.form-en} we provide useful expressions for the polaron energy. Section~\ref{sec.wfc}
describes how to visualize the polaron wavefunctions, and Sec.~\ref{sec.dft-lp} established the
formal link between the present approach and the Landau-Pekar theory described in Sec.~\ref{sec.lp}.
In Sec.~\ref{sec.sic} we discuss the SIC employed in this work, and how
it relates to the polaron equations derived in Sec.~\ref{sec.pol-eqs}. The technical details of
our implementation and the computational setup for the calculations are described in Sec.~\ref{sec.setup}.
In particular we give details of all DFT calculations (Sec.~\ref{sec.dft}), of the nonlinear
eigenvalue solver (Sec.~\ref{sec.solvers}), and basic information on each of the compounds 
considered (Sec.~\ref{sec.systems}). In Sec.~\ref{sec.results} we illustrate our results. 
First we validate our SIC against previous work using $\alpha$-quartz as a test case 
(Sec.~\ref{sec.validation}). Then we discuss the dependence of the polaron energies on supercell
size and compare with previous work and SIC calculations in Sec.~\ref{sec.convergence}.
We show polaron wavefunctions and lattice distortions in Sec.~\ref{sec.wavefunction},
and we compare our results with explicit supercell calculations. The spectral decomposition of 
the polaron into KS states and normal modes is presented in Sec.~\ref{sec.spectral}.
 In Sec.~\ref{sec.future} we discuss possible future work to link the present formalism with 
advanced many-body approaches for model Hamiltonians, and in Sec.~\ref{sec.conclusions} we 
draw our conclusions.

\section{The Landau-Pekar model}\label{sec.lp}

In this section we summarize the original derivation of the Landau-Pekar (LP) model,\cite{Landau1933,Pekar1946}
since this model provides a very useful starting point to understand our {\it ab initio} approach 
described in Sec.~\ref{sec.abinitio}. 

The LP model is a simple yet powerful framework for studying a single electron added to a polar insulator. 
The key assumption of this model is that the electron wavefunction extends over spatial dimensions spanning 
many crystal unit cells. As a consequence, the atomistic details of the crystal are neglected; the 
interaction of the added electron with the valence manifold is described via the effective-mass approximation 
and thus only enters the kinetic energy; the interaction of this electron with the ionic lattice is described 
via a continuum electrostatic model. 
The total energy of the LP model is written as:
  \begin{equation}
  \label{eq:LP.1}
  E_{\rm LP} = \frac{\hbar^2}{2m^*} \!\int\! d\br\, |\nabla \psi|^2 
  + \frac{1}{2}\int \!d\br\,\, \bE\cdot\bD,
  \end{equation}
where $\psi(\br)$ is the wavefunction of the added electron, $\bE(\br)$ is the self-consistent electric
field and $\bD(\br)$ is the electric displacement field. The first term on the r.h.s.\ of Eq.~(\ref{eq:LP.1}) 
represents the band energy of the extra electron, and includes electron-electron interactions via the 
conduction band effective mass $m^*$. The second term represents the total electrostatic energy of the 
dielectric.\cite{Jackson1998}

The displacement field $\bD$ is related to the density of free carriers, and therefore to the 
wavefunction of the excess electron, by the relation $\nabla\cdot \bD = -e |\psi(\br)|^2$ ($e$ is
the electron charge), or equivalently:
  \begin{equation}
  \label{eq:LP.2}
  \bD = \frac{e}{4\pi}\nabla \!\int \!d\brp\, \frac{\,\,|\psi(\brp)|^2}{|\br-\brp|}.
  \end{equation}
The displacement field is also related to the self-consistent electric field via $\bD = \e_0 \e^0 \bE$, where
$\e_0$ is the vacuum permittivity and $\e^0$ is the static dielectric constant. By replacing
Eq.~(\ref{eq:LP.2}) into (\ref{eq:LP.1}) we obtain the total electrostatic energy:
  \begin{equation}
  \label{eq:LP.3}
  \frac{1}{2}\int \!d\br\,\, \bE\cdot\bD =
       \frac{1}{2} \frac{e^2}{4\pi\e_0} \frac{1}{\e^0}\!\int \!\!d\br\, d\brp
       \frac{\,|\psi(\br)|^2|\psi(\brp)|^2}{|\br-\brp|}.
  \end{equation}
In this expression the electric field $\bE$ includes contributions from both the electronic screening 
and the lattice screening. Since the electronic screening energy is already accounted for in the band 
structure term in Eq.~(\ref{eq:LP.1}), we need to subtract this contribution from Eq.~(\ref{eq:LP.3}).
The electronic-only contribution is simply obtained by evaluating Eq.~(\ref{eq:LP.3}) with the ionic 
screening turned off, i.e.\ by using the high-frequency (electronic) permittivity $\e^\infty$ instead 
of the static (electronic and ionic) permittivity $\e^0$. After removing this contribution the 
electrostatic energy reads:
  \begin{equation}
  \label{eq:LP.4}
  \frac{1}{2}\!\int \!d\br\, \bE\cdot\bD = \frac{1}{2} \frac{e^2}{4\pi\e_0}\! 
  \left(\frac{1}{\e^0}-\frac{1}{\e^\infty}\right)\!\!\int \!\!d\br\, d\brp \frac{\, 
  |\psi(\br)|^2|\psi(\brp)|^2}{|\br-\brp|}.
  \end{equation}
By defining $1/\kappa = 1/\e^\infty - 1/\e^0$,\cite{Alexandrov2010} we can rewrite Eq.~(\ref{eq:LP.1}) 
as a functional of the polaron wavefunction:
  \begin{eqnarray}
  \label{eq:LP.5}
  E_{\rm LP}[\psi] &=& \frac{\hbar^2}{2m^*} \!\int\! d\br\, |\nabla \psi(\br)|^2 
  \nonumber \\ &-& \frac{1}{2} \frac{e^2}{4\pi\e_0} \frac{1}{\kappa}\int \!\!d\br\, 
  d\brp \frac{\,|\psi(\br)|^2|\psi(\brp)|^2}{|\br-\brp|}.
  \end{eqnarray}
The ground-state energy of the LP polaron is found by minimizing this functional with respect 
to $\psi$, subject to the constraint provided by the normalization condition $\int \!d\br\,
|\psi(\br)|^2 =1$. This problem can be solved by transforming it into an unconstrained 
minimization with the normalization incorporated via the Lagrange multiplier $\ve$:
  \begin{eqnarray} \label{eq:LP.6}
  E_{\rm LP}^{\prime}[\psi,\ve] &=& \frac{\hbar^2}{2m^*} \!\int\! d\br\, |\nabla \psi(\br)|^2 
  \nonumber \\ &-&\frac{1}{2}\frac{e^2}{4\pi\e_0}\frac{1}{\kappa}
  \int \! d\br \, d\brp \frac{\,|\psi(\br)|^2\,|\psi(\brp)|^2}{|\br-\brp|}
  \nonumber \\ &-& \ve\left( \int \!d\br\, |\psi(\br)|^2 - 1\right).
  \end{eqnarray}
By setting to zero the two functional derivatives $\d E_{\rm LP}^\prime /\d \psi^*$ and 
$\d E_{\rm LP}^\prime /\d \ve$ one obtains a Schr\"odinger-type eigenvalue problem for $\psi$:
  \begin{eqnarray} \label{eq:LP.7}
  &&-\frac{\hbar^2}{2m^*} \nabla^2 \psi(\br) - \frac{e^2}{4\pi\e_0} \frac{1}{\kappa}\int \! d\brp 
  \frac{\,\,|\psi(\brp)|^2}{|\br-\brp|} \psi(\br) = \ve\,\psi(\br),\hspace{10pt}  \\
  &&\int \!d\br\, |\psi(\br)|^2 =1.
  \end{eqnarray}
Here the eigenvalue $\ve$ carries the meaning of an energy, but it is not the total energy of 
the polaron. This is seen by projecting Eq.~(\ref{eq:LP.7}) onto $\psi^*$ and comparing with 
Eq.~(\ref{eq:LP.5}):
  \begin{equation} \label{eq:LP.8}
  E_{\rm LP} =  \ve + \frac{1}{2}\frac{e^2}{4\pi\e_0}\frac{1}{\kappa}
  \int \! d\br \, d\brp \frac{\,|\psi(\br)|^2\,|\psi(\brp)|^2}{|\br-\brp|}.
  \end{equation}
Equation~(\ref{eq:LP.7}) provides an intuitive understanding of the nature of polaron self-trapping 
in the LP model. Let us consider for example a normalized trial wavefunction $\psi(\br) = 
(\pi r_{\rm p}^3)^{-1/2} \exp(-|\br|/r_{\rm p})$.\cite{Alexandrov2010} Using this
trial function, it is evident that that minimization of the kinetic energy term in Eq.~(\ref{eq:LP.7}) 
favors delocalization (larger $r_{\rm p}$), while the minimization of the Coulomb term favors 
localization (smaller $r_{\rm p}$). The polaron size $r_{\rm p}$ results from a trade-off between 
these competing effects. By replacing the above exponential ansatz in Eq.~(\ref{eq:LP.5}), one obtains 
a simple estimate for the energy as a function of the polaron size $r_{\rm p}$:
  \begin{equation}
  \label{eq:LP.11}
  E_{\rm LP}(r_{\rm p}) = \frac{\hbar^2}{2m^* r_{\rm p}^2} 
  - \frac{5}{16}\frac{1}{\kappa}\frac{e^2}{4\pi\e_0 r_{\rm p}}.
  \end{equation}
The minimum of this function is given by:
  \begin{equation}
  \label{eq:LP.12}
  \frac{r_{\rm p,min}}{a_0} = \frac{16}{5} \frac{\kappa}{m^*\!/m_e},
  \end{equation}
where $a_0$ denotes the Bohr radius and $m_e$ is the free electron mass. By replacing Eq.~(\ref{eq:LP.12})
inside Eq.~(\ref{eq:LP.11}) we find the ground-state energy:
  \begin{equation}
  \label{eq:LP.13}
  \frac{E_{\rm LP,min}}{E_{\rm Ha}} = -\frac{25}{512} \frac{m^*/m_e}{\kappa^2},
  \end{equation}
with $E_{\rm Ha}$ being the Hartree energy.
Furthermore, by using Eq.~(\ref{eq:LP.8}) we obtain the polaron eigenvalue:\cite{Alexandrov2010}
  \begin{equation}
  \label{eq:LP.13b}
  \frac{\ve}{E_{\rm Ha}} = -\frac{75}{512} \frac{m^*/m_e}{\kappa^2}.
  \end{equation}
The energy given by Eq.~(\ref{eq:LP.13}) can also be expressed by using the standard polaron coupling
constant $\alpha$, which is defined as:
  \begin{equation}
  \label{eq:LP.14}
  \a = \frac{e^2}{4\pi\e_0\hbar}\sqrt{\frac{m^*}{2\hbar\omega_{\text{LO}}}}\frac{1}{\k},
  \end{equation}
where $\omega_{\text{LO}}$ is the characteristic frequency of longitudinal optical phonons. By combining 
Eqs.~(\ref{eq:LP.13}) and (\ref{eq:LP.14}) one obtains the standard result:
  \begin{equation}
  \label{eq:LP.15}
  E_{\rm LP,min} = -\frac{50}{512} \, \a^2\, \hbar\omega_{\text{LO}},
  \end{equation}
which is very close to the original variational solution by Pekar.\cite{Pekar1946}
Much work has been done to improve on the simple exponential ansatz employed in this brief overview
of the LP model. However, apart from obtaining a more accurate prefactor in front of the term 
$\a^2 \, \hbar\w_{\text{LO}}$ in Eq.~(\ref{eq:LP.15}), these improvements do not change the qualitative features 
of the solution. This is a consequence of the fact that Eq.~(\ref{eq:LP.5}) can be written in a 
scale-invariant form by defining $\psi(\br) = a^{-3/2}\phi (\br/a)$ with $a = \k \,a_0 / (m^*/m_e)$, so that:
  \begin{equation}
  \label{eq:LP.16}
  \frac{E_{\rm LP}}{\a^2\,\hbar \w} = \int\!d\br\, |\nabla \phi(\br)|^2 
  - \int \frac{|\phi(\br)|^2|\phi(\brp)|^2}{|\br-\brp|}d\br \,d\brp, 
  \end{equation}
subject to the normalization condition \mbox{$\int \!d\br\, |\phi(\br)|^2 = 1$}. The direct numerical 
solution\cite{Miyake1975} of Eq.~(\ref{eq:LP.16}) yields a wavefunction which 
is very close to the original variational result found by Pekar using a modified 
exponential.\cite{Pekar1946,Alexandrov2010,Devreese2009,Miyake1975}

From Eq.~(\ref{eq:LP.13}) we see that the formation of a localized polaron is only possible when
$\e^0 > \e^\infty$, that is in polar crystals. This leaves out those polarons that can form in
non-polar semiconductors. In addition, since the electron-phonon coupling mechanism is related to
the ionic dielectric response, i.e. to the long-range Fr\"ohlich potential generated by lattice 
distortions, the LP model also leaves out acoustic and piezo-acoustic polarons. Further limitations
of the model are that it assumes an isotropic dielectric, and does not take into account the atomistic
nature of the crystal lattice. In Ref.~\onlinecite{Devreese2009}
it was pointed out that the LP model is essentially never valid, because it relies on the assumption 
of large polarons in order to use continuum electrostatics, but its results tend to be accurate 
in the regime of strong coupling, that is for small polarons, in contrast with the starting 
hypothesis. The LP model is said to describe strong-coupling polarons because the energy given by
Eq.~(\ref{eq:LP.15}) is almost the same as that obtained in the strong-coupling limit of the Feynman 
theory, $-\alpha^2\,\hbar\w/3\pi$.\cite{Feynman1955}

In the following section we show how the essential physics of the LP model can be 
retained by moving to an {\it ab initio} formalism based on density-functional theory, and that 
most of the intrinsic limitations of the model can be overcome in this new framework.

\section{Polarons in density-functional theory}\label{sec.abinitio}

\subsection{Derivation of the polaron equations}\label{sec.pol-eqs}

In order to develop an {\it ab initio} theory of polarons, we take the view that standard density-functional 
theory (DFT) implementations contain most of the essential physics, and can serve as a useful starting
point. The modification to remove the self-interaction error in standard DFT will be discussed in 
Sec.~\ref{sec.sic}.

DFT already incorporates the physics of the LP model: if we add an electron in an otherwise 
empty conduction band of a semiconductor or insulator, the ions experience an additional force that causes 
them to screen the extra charge. This notion is well established, and has been exploited in several 
investigations of {\it small} polarons, i.e. polarons with a spatial extension corresponding to one 
or few atomic orbitals.\cite{Mauri2005,Feng2013,kokott2018}

The main limitation of such direct calculations is that only small polarons can be investigated, because
intermediate-size and large polarons would require prohibitively time-consuming calculations with supercells 
containing many thousands of atoms. Another limitation is that with direct calculations it is not 
possible to analyze the individual contributions to the polaron formation, for example which phonons 
are responsible for the self-trapping, and which electrons participate in the polaron wavefunction.
Lastly, direct calculations are very sensitive to the choice of the DFT exchange and correlation 
functional, mostly due to the self interaction error, making it very challenging to obtain reliable polaron formation energies.

To overcome these limitations it is desirable to formulate an {\it ab initio} theory of polarons which
does not require large supercell calculations, and where the individual contributions to the polaron
energy and wavefunctions are easily recognizable. In the following we propose a new framework to
address these challenges.

We start by writing the DFT total energy of a semiconducting or insulating crystal, with the valence bands 
fully occupied and the conduction bands empty. We consider a Born-von Karman supercell of the crystal, 
containing $N_p$ unit cells of volume $\Omega$. We follow the notation of Ref.~\onlinecite{Giustino2017} 
and use $\btau_{\k p}$ and $\tau_{\k p \a}$ to indicate the position and Cartesian coordinates of the atom 
$\k$ in the unit cell $p$ along the Cartesian direction $\a$, respectively. The atom $\k$ has a charge 
$e\,Z_\k$; we shall write the equations with an all-electron implementation in mind; the transposition to
a pseudopotential formalism is obvious. The KS states have wavefunctions $\psi_{n\bk}(\br)$ with energies $\ve_{n\bk}$,
where $n$ is the band index and $\bk$ the wavevector. The wavefunctions 
are normalized in the supercell, and we have $N_p$ $\bk$-points on a uniform grid. With this notation 
the electron density reads $n = n^\up+n^\down$, with $n^\up(\br) = n^\down(\br) = 
\sum_{v\bk} |\psi_{v\bk}(\br)|^2$ and the subscript $v$ running over all occupied states. The system is 
assumed to be spin-degenerate in the ground state. The DFT total energy of the entire supercell reads:
  \begin{eqnarray}
  \label{eq:DFT.1}
  &&\frac{E[\{\psi_{v\bk}\},\{\btau_{\k \a}\}]}{E_{\rm Ha}} =
  - 2 \sum_{v\bk} \int d\br \,\psi_{v\bk}^* \frac{a_0^2\nabla^2}{2} \psi_{v\bk} \nonumber \\
  && +\frac{1}{2} \sum_\bT \int d\br d\brp \frac{a_0\,n(\br)n(\brp)}{|\br-\brp-\bT|}
  +\frac{E_{xc}[n^\up,n^\down]}{E_{\rm Ha}} \nonumber \\
  &&- \sum_{\k p \bT} \int d\br \frac{a_0\, Z_\k n(\br)}{|\br-\btau_{\k p} - \bT|}
  +\frac{1}{2} \sum_{\substack{\k p \bT\\\k' p'}} \frac{a_0\,Z_\k Z_{\k'}}{|\btau_{\k p} -
  \btau_{\k' p'}-\bT|}, \nonumber \\
  \end{eqnarray}
where $\bT$ is a vector of the supercell lattice, and all integrals are evaluated over the supercell.
In the last term the contribution from $\k p = \k' p'$ is omitted when $\bT=0$.
We now call $\btau^0_{\k p}$ the atomic positions at equilibrium in the ground state, so that a
general ionic coordinate reads $\btau_{\k p} = \btau^0_{\k p} + \Delta \btau_{\k p}$. Similarly, we call
$\psi_{v\bk}^0$ the wavefunctions obtained with the atoms in the equilibrium positions, and $n^0$ the
corresponding density. To second order in the displacements $\Delta \btau_{\k p}$, the total energy 
in Eq.~(\ref{eq:DFT.1}) can be written as:
  \begin{eqnarray}
  \label{eq:DFT.2}
  && E[\{\psi_{v\bk}\},\{\btau_{\k p}\}] = E[\{\psi^0_{v\bk}\},\{\btau^0_{\k p}\}] \nonumber \\ && 
  \hspace{20pt}+
  \frac{1}{2}\sum_{\substack{\kappa\a p\\\kappa'\a' p'}} C^0_{\k \a p, \k'\a' p'} 
  \Delta\tau_{\k \a p} \Delta\tau_{\k' \a' p'} + \mathcal{O}(\Delta\tau^3),\hspace{20pt}
  \end{eqnarray}
where $C^0_{\k \a p, \k'\a' p'}$ is the usual matrix of interatomic force constants,\cite{Baroni2001,
Gonze1997,Giustino2017} evaluated for the ground state. Upon adding an extra electron to the ground state, 
we fill one conduction state and the system becomes spin polarized. Before proceeding we emphasize
that the same reasoning can be made for the case of a hole at the top of the valence bands; the
formalism is entirely symmetric in this respect. Let us call the wavefunction of the excess electron 
$\psi$, and its associated density $\Delta n = |\psi|^2$. For definiteness we say that this 
extra electron carries a spin up. We also add a compensating jellium background, $-1/N_p\Omega$, to 
avoid the Coulomb divergence. The total energy from Eq.~(\ref{eq:DFT.1}) is modified as follows:
  \begin{eqnarray}
  \label{eq:DFT.3}
    &&\hspace{-12pt}\frac{E[\psi,\{\psi_{v\bk}\},\{\btau_{\k p}\}]}{E_{\rm Ha}} = 
    - 2 \sum_{v\bk} \int d\br \,\psi_{v\bk}^*\frac{a_0^2\nabla^2}{2} \psi_{v\bk} \nonumber \\
    && - \int d\br \,\psi^* \frac{a_0^2\nabla^2}{2} \psi 
     + \frac{E_{xc}[n^\uparrow+\Delta n,n^\down]}{E_{\rm Ha}} \nonumber \\
    && + \frac{1}{2} \sum_\bT \int d\br d\brp \frac{a_0}{|\br-\brp-\bT|} \nonumber \\
    && \times [n(\br)+\Delta n(\br)-1/N_p\Omega] [n(\brp)+\Delta n(\brp)-1/N_p\Omega]\nonumber \\[6pt]
    && - \sum_{\k p \bT} \int d\br \frac{a_0\,Z_\k [n(\br)+\Delta n(\br)-1/N_p\Omega]}
    {|\br-\btau_{\k p} - \bT|} \nonumber \\ && + \frac{1}{2} \sum_{\k p,\k' p' \bT} 
    \frac{a_0\, Z_\k Z_{\k'}}{|\btau_{\k p} - \btau_{\k' p'}-\bT|}.
  \end{eqnarray}
In order to proceed we make the following key observation: the addition of a single electron to a system
of many electrons will modify the electron density only slightly. Indeed, in the limit of very large
polaron the extra electron density at any point will be of the order of $(N_p\Omega)^{-1} \ll n$; in the
limit of very small polaron the density will be of the order of $\Omega^{-1}$ in one unit cell, and negligible
in the others. Following this argument, in the following we make the approximations that, upon adding one
electron, $\Delta n \ll n$ almost everywhere, and as a result the valence wavefunctions $\psi_{v\bk}$ 
remain unaltered. The latter approximation allows us to expand the exchange and correlation energy as follows:
  \begin{eqnarray}
  \label{eq:DFT.4}
    && \hspace{-15pt}E_{xc}[n^\uparrow+\Delta n,n^\down]  = E_{xc}[n^\uparrow,n^\down] + \int d\br 
    \frac{\d E_{xc}}{\d n^\up}\Delta n(\br) \nonumber \\ && + \int d\br d\brp \frac{1}{2}\frac{\d^2 E_{xc}}
    {\d n^\up \d n^\up} \Delta n(\br)\Delta n(\brp) + \mathcal{O}(\Delta n^3).
  \end{eqnarray}
By combining Eqs.~(\ref{eq:DFT.1})-(\ref{eq:DFT.4}) and rearranging we find:
  \begin{eqnarray}
  \label{eq:DFT.5}
    &&E[\psi,\{\psi_{v\bk}\},\{\btau_{\k p}\}] = E[\{\psi^0_{v\bk}\},\{\btau^0_{\k p}\}]+
      \nonumber \\
    && \qquad + \frac{1}{2}\sum_{\substack{\kappa\a p\\\kappa'\a' p'}} C^0_{\k \a p, \k'\a' p'} 
    \Delta\tau_{\k \a p} \Delta\tau_{\k' \a' p'} \nonumber \\[-8pt]
    &&\qquad+ E_{\rm Ha} \int \! d\br\, \psi^*(\br) \!\Bigg[ -\frac{a_0^2}{2} \nabla^2 
    +\!\sum_\bT \!\int\! d\brp \frac{a_0 n(\brp)}{|\br-\brp-\bT|} \nonumber \\
    && \qquad \left.
    -\sum_{\k p \bT} \frac{Z_\k a_0}{|\br-\btau_{\k p} - \bT|} 
    + \frac{1}{E_{\rm Ha}}\frac{\d E_{xc}}{\d n^\up} \right]\! \psi(\br) \nonumber \\
    &&\qquad+\frac{1}{2}E_{\rm Ha}\Bigg[ 
    \int d\br d\brp \frac{1}{E_{\rm Ha}}\frac{\d^2 E_{xc}}{\d n^\up \d n^\up} \Delta n(\br)\Delta n(\brp)
    \nonumber \\
    &&\qquad \left. +\sum_\bT \int d\br d\brp \frac{
    [\Delta n(\br)-1/N_p\Omega][\Delta n(\brp)-1/N_p\Omega]}{|\br-\brp-\bT|/a_0}
    \right] \nonumber \\ &&\qquad + E_{\rm B} + \mathcal{O}(\Delta\tau^3) + \mathcal{O}(\Delta n^3),
  \end{eqnarray}
where $E_{\rm B}$ is a constant term arising from the jellium background. Inside the square brackets 
in the third and fourth lines of this equation we recognize the KS Hamiltonian 
$\hH_{\rm KS}[n(\br),\{\btau_{\k p}\}]$ associated
with the occupied manifold in {\it absence} of the excess electron. In analogy with Eq.~(\ref{eq:DFT.2}),
we can rewrite this term by performing a Taylor expansion around the equilibrium atomic coordinates:
  \begin{eqnarray}
  \label{eq:DFT.6}
  &&\hspace{-15pt}\hH_{\rm KS}[n(\br),\{\btau_{\k p}\}] = \hH_{\rm KS}[n^0(\br),\{\btau^0_{\k p}\}]
  \nonumber \\ &&+\sum_{\k \a p} \frac{\D V_{\rm KS}^0 }{\D \tau_{\k \a p}} \Delta \tau_{\k \a p} 
  +\mathcal{O}(\Delta \tau^2),
  \end{eqnarray}
where we use $V_{\rm KS}^0$ to indicate the KS self-consistent potential at equilibrium, in the
absence of the excess electron. To keep the formalism as simple as possible, we truncate the expansion 
to first order in $\Delta \tau_{\k \a p}$. This is the lowest order that admits non-trivial
solutions, that is self-trapped polarons.

The fifth and sixth lines of Eq.~(\ref{eq:DFT.5}) contain the Hartree, exchange, and correlation 
{\it self-interaction}
of the excess electron. These are spurious contributions which artificially increase the energy needed
to form a polaron, and which tend to delocalize the polaron wavefunctions. For the time being we neglect
these terms. In Sec.~\ref{sec.sic} we show that the correct procedure to deal with these terms is to modify 
the exchange-correlation functional $E_{xc}$ by including suitable SICs. The 
resulting formalism is robust and mathematically elegant (validation tests are presented in 
Sec.~\ref{sec.validation}).

Now we can combine Eqs.~(\ref{eq:DFT.5}) and (\ref{eq:DFT.6}) to obtain our final expression for the
DFT functional of a polaron. At this point we omit the fifth, sixth, and seventh lines of Eq.~(\ref{eq:DFT.5}),
and we use the short-hand notation $\hH_{\rm KS}^0$ for $\hH_{\rm KS}[n^0(\br),\{\btau^0_{\k p}\}]$:
  \begin{eqnarray}
  \label{eq:DFT.7}
  && \hspace{-15pt}E_{\rm p}[\psi,\{\Delta\tau_{\k \a p}\}] 
  = E[\{\psi^0_{v\bk}\},\{\btau^0_{\k p}\}] \nonumber \\[2pt]
  && +
  \frac{1}{2}\sum_{\substack{\k\a p\\\k'\a' p'}} C^0_{\k \a p, \k'\a' p'} 
  \Delta\tau_{\k \a p} \Delta\tau_{\k' \a' p'} \nonumber \\[-2pt] &&
  +\int \! d\br\, \psi^*(\br) \!\left[ \hH_{\rm KS}^0 + 
  \sum_{\k \a p} \frac{\D V_{\rm KS}^0 }{\D \tau_{\k \a p}} \Delta \tau_{\k \a p} \right]\! \psi(\br).
  \end{eqnarray}
The functional $E_{\rm p}[\psi,\{\Delta\tau_{\k \a p}\}]$ defined by this equation constitutes
the DFT counterpart of the Laundau-Pekar functional in Eq.~(\ref{eq:LP.8}). Also in this case we can
take into account the normalization constraint on the wavefunction by introducing the Lagrange multiplier 
$\ve$. By setting to zero the derivatives with respect to $\psi^*$ and $\Delta\tau_{\k \a p}$, we 
find the coupled system of equations:
  \begin{eqnarray}
  \label{eq:DFT.8}
  &&\hspace{-18pt}\displaystyle \,\,\frac{\d }{\d \psi^*}\left[ E_{\rm p} -\ve \Big(\int\!d\br\, 
  |\psi(\br)|^2-1\Big)\right] = 0 \,:  \nonumber \\ && \displaystyle \hH_{\rm KS}^0 
  \psi(\br) + \sum_{\k \a p} \frac{\D V_{\rm KS}^0 }{\D \tau_{\k \a p}} \Delta \tau_{\k \a p} 
  \psi(\br) = \ve \, \psi(\br), \\[-3pt]
  \label{eq:DFT.9}
  &&\hspace{-18pt}\displaystyle \,\,\frac{\d E_{\rm p}}{\d \Delta\tau_{\k \a p}} = 0 \,:  \nonumber\\[-3pt]
  && \displaystyle \Delta\tau_{\k \a p} = 
  -\!\!\!\sum_{\k' \a' p'} \!\!(C^0)^{-1}_{\k \a p, \k' \a' p'} \!\!\int \!\!d\br\,
  \frac{\D V_{\rm KS}^0 }{\D \tau_{\k' \a' p'}}|\psi(\br)|^2 \!.
  \end{eqnarray}
This coupled system of equations defines a self-consistent problem in $\psi$ and $\Delta\tau_{\k \a p}$, 
whose solution yields the polaron wavefunction and the associated pattern of atomic displacements. In 
order to emphasize the analogy with the Landau-Pekar polaron discussed in Sec.~\ref{sec.lp}, it is 
convenient to replace Eq.~(\ref{eq:DFT.9}) inside (\ref{eq:DFT.8}). The result is:
  \begin{equation}
  \label{eq:DFT.10}
  \hH_{\rm KS}^0 \psi(\br) -\! \int \!\!d\brp\, K^0(\br,\brp)\,|\psi(\brp)|^2\, \psi(\br) = \ve \, \psi(\br),
  \end{equation}
having defined the `polaron kernel' $K^0(\br,\brp)$ as:
  \begin{equation}
  \label{eq:DFT.11}
  K^0(\br,\brp) = \sum_{\k \a p}\sum_{\k' \a' p'} \frac{\D V_{\rm KS}^0 (\br)}{\D \tau_{\k \a p}} 
  (C^0)^{-1}_{\k \a p, \k' \a' p'} \frac{\D V_{\rm KS}^0(\brp) }{\D \tau_{\k' \a' p'}}.
  \end{equation}
In this form the similarity with Eq.~(\ref{eq:LP.7}) is evident: the KS Hamiltonian in
Eq.~(\ref{eq:DFT.10}) is the counterpart of the kinetic energy with the band effective mass in
the LP model, while the kernel is the counterpart of the self-trapping potential. We will
elaborate on this analogy in Sec.~\ref{sec.dft-lp}.

\subsection{The formation energy of a polaron and the meaning of the polaron eigenvalue}\label{sec.pol-en}

As in the case of the LP model, the eigenvalue $\ve$ appearing in Eq.~(\ref{eq:DFT.10}) does not
correspond to the energy of the polaron. To see this it is convenient to define the polaron 
{\it formation energy} $\Delta E_f$ as the energy required to trap a conduction band state into 
a localized polaron:
  \begin{eqnarray}
  \label{eq:formation.1}
  &&\Delta E_f = {\rm min}\, E_{\rm p}[\psi,\{\Delta\tau_{\k \a p}\}] - {\rm min}\,
  E_{\rm p}[\psi,\{\Delta\tau_{\k \a p}=0\}]. \nonumber \\[2pt]
  \end{eqnarray}
Here $E_{\rm p}$ is the functional defined by Eq.~(\ref{eq:DFT.7}). This definition yields the energy
gained by the system when a delocalized conduction electron becomes self-trapped, and allows us to
separate the energetics of the polaron formation from that of the electron addition into the conduction
band of the insulator/semiconductor with the ions in the equilibrium positions. 
By using Eqs.~(\ref{eq:DFT.7}), (\ref{eq:DFT.9}) and (\ref{eq:DFT.11}) in this expression we find:
  \begin{eqnarray}
  \label{eq:formation.2}
  \Delta E_f &=& 
  \int \! d\br\, \psi^*(\br) \!\left( \hH_{\rm KS}^0 - \ve_{\rm CBM} \right)\! \psi(\br)
  \nonumber \\  &-&\frac{1}{2}\int \! d\br\,d\brp \,|\psi(\br)|^2 K^0(\br,\brp) |\psi(\brp)|^2,
  \end{eqnarray}
where $\ve_{\rm CBM}$ is the KS eigenvalue of the conduction band bottom.
Similarly, we can obtain an expression for the Lagrange multiplier $\ve$ in Eq.~(\ref{eq:DFT.10})
by projecting onto $\psi^*$:
  \begin{eqnarray}
  \label{eq:formation.3}
  \ve - \ve_{\rm CBM} &=& 
  \int \! d\br\, \psi^*(\br) \!\left( \hH_{\rm KS}^0 - \ve_{\rm CBM} \right)\! \psi(\br) \nonumber \\
  &-&\!\int \! d\br\,d\brp \,|\psi(\br)|^2 K^0(\br,\brp) |\psi(\brp)|^2.
  \end{eqnarray} 
By subtracting the last two equations we obtain a simple relation between the formation energy 
$\Delta E_f$ and the eigenvalue $\ve$:
  \begin{equation}
  \label{eq:formation.4}
  \Delta E_f = \ve - \ve_{\rm CBM}
  +\frac{1}{2}\int \! d\br\,d\brp \,|\psi(\br)|^2 K^0(\br,\brp) |\psi(\brp)|^2.
  \end{equation} 
This result shows that the Lagrange multiplier contains a double counting of the Coulomb energy, which has
to be removed in order to obtain the formation energy. This is analogous to the relation
between the DFT total energy and the sum of the band eigenvalues.\cite{Giustino2014}

By using Eqs.~(\ref{eq:DFT.9}) and (\ref{eq:DFT.11}) we can rewrite Eq.~(\ref{eq:formation.4})
as follows:
  \begin{equation}
  \label{eq:formation.5}
  \Delta E_f = \ve - \ve_{\rm CBM}
  +\frac{1}{2}\sum_{\substack{\kappa\a p\\\kappa'\a' p'}} C^0_{\k \a p, \k'\a' p'} 
  \Delta\tau_{\k \a p} \Delta\tau_{\k' \a' p'}.
  \end{equation}
This expression for the formation energy can be interpreted in the context of Franck-Condon
principle: the difference $\ve_{\rm CBM}-\ve$ can be thought of as the energy required
for an ultrafast excitation to promote the electron from the polaron state to a band
state at the bottom of the conduction manifold, while the ions are still in 
the distorted polaron state; the sum on the r.h.s.\ then corresponds to energy released 
by the distorted lattice upon relaxation. The same interpretation is often discussed
in relation to the LP model.\citep{Devreese2009}

\subsection{Polaron equations in the basis of Kohn-Sham states and phonon modes}\label{sec.pol-eqs-ks}

For practical {\it ab initio} calculations it is convenient to recast the equations derived in 
Sec.~\ref{sec.pol-en} in a reciprocal space formulation. Since the KS states in the ground 
state form a complete basis, we can expand the polaron wavefunction as:
  \begin{equation}
  \label{eq:basis.1}
  \psi(\br) = \frac{1}{\sqrt{N_p}}\sum_{n\bk} A_{n\bk} \psi_{n\bk}(\br), 
  \end{equation}
where the summation is restricted to the unoccupied (conduction) states since we are assuming that the
valence band manifold remains unchanged. From the normalization of the KS states $\psi_{n\bk}$ 
and the polaron wavefunction $\psi$ it follows:
  \begin{equation}
  \label{eq:basis.1b}
  \frac{1}{N_p}\sum_{n\bk} |A_{n\bk}|^2 = 1.
  \end{equation}
Now we replace Eq.~(\ref{eq:basis.1}) inside Eq.~(\ref{eq:DFT.10}) and project both sides on a KS
state. To carry out the algebra it is useful to keep in mind the standard relations between the 
electron-phonon matrix elements, the interatomic force constants, and the vibrational 
eigenmodes:\cite{Giustino2017}
  \begin{eqnarray}
  \label{eq:basis.2}
  g_{mn\nu}(\bk,\bq) &=& \sum_{\k\a p} 
  \left(\frac{\hbar}{2M_\k \w_{\bq\nu}}\right)^{\!\!1/2}\!\!\!  e_{\k\a,\nu}(\bq)\, 
  e^{i\bq\cdot \bR_p} \nonumber \\ &\times&
  \int \!d\br \,\psi^*_{m\bk+\bq}(\br) \frac{\D V_{\rm KS}^0 (\br)}{\D \tau_{\k \a p}} \,\psi_{n\bk}(\br),
  \end{eqnarray}
  \begin{equation}
  \label{eq:basis.3}
  (C^0)^{-1}_{\k \a p, \k' \a' p'}  = \frac{1}{N_p}\sum_{\bq\nu} 
  \frac{e_{\k\a,\nu}(\bq) e^*_{\k'\a',\nu}(\bq)}{\sqrt{M_\k M_{\k'}}\w_{\bq\nu}^2}
  e^{i\bq\cdot (\bR_p-\bR_{p'})}.
  \end{equation}
Here $e_{\k\a,\nu}(\bq)$ denotes orthonormal vibrational modes for the wavevector $\bq$ and branch $\nu$,
with frequency $\w_{\bq\nu}$. $M_\k$ is the mass of the $\k$-atom, $\bR_p$ is a vector of the direct lattice
of the crystal unit cell. The integral is over the supercell, and $g_{mn\nu}(\bk,\bq)$ is the matrix
element for the scattering of an electron $\psi_{n\bk}$ into $\psi_{m\bk+\bq}$ via the phonon $\bq\nu$;
it has dimensions of an energy. By combining Eqs.~(\ref{eq:DFT.10})-(\ref{eq:DFT.11}) and 
(\ref{eq:basis.1})-(\ref{eq:basis.3}) we arrive at the self-consistent eigenvalue problem:
  \begin{eqnarray}
  \label{eq:basis.4}
  && \frac{2}{N_p}\sum_{\bq m\nu}  
  B_{\bq\nu}\,g^*_{mn\nu}(\bk,\bq) \,A_{m\bk+\bq} =
  (\ve_{n\bk}-\ve)\,A_{n\bk}, \hspace{10pt}\\
  \label{eq:basis.5}
  &&  B_{\bq\nu} = \frac{1}{N_p} \sum_{mn\bk} A^*_{m\bk+\bq}
  \,\frac{g_{mn\nu}(\bk,\bq)}{\hbar\w_{\bq\nu}}\,A_{n\bk} .
  \end{eqnarray}
The operator on the left-hand side of Eq.~(\ref{eq:basis.4}) is Hermitian. 
This can be verified after noting that from Eq.~(\ref{eq:basis.5}) we have 
$B^*_{\bq\nu}=B_{-\bq+\bG,\nu}$, where $\bG$ is a reciprocal lattice vector that folds $-\bq$ back 
into the first Brillouin zone (possibly $\bG=0$). The periodicity of $B_{\bq \nu}$ is inherited 
from the choice of a periodic gauge for both the KS states and phonon modes.
Furthermore, by taking the complex conjugate of Eq.~(\ref{eq:basis.4}) and using
$g^*_{mn\nu}(\bk,\bq) = g_{mn\nu}(-\bk+\bG,-\bq)$ from time-reversal symmetry,
it can be seen that if $A_{n\bk}$ is a solution vector, then also $A^*_{n,-\bk+\bG}$ is a solution for the
same eigenvalue. This implies that, apart from a non-essential phase, $A_{n,-\bk+\bG}=A^*_{n\bk}$. 
By using this property in the expansion Eq.~(\ref{eq:basis.1}) we see that the polaron wavefunction
$\psi$ has to be real-valued. 

Equations~(\ref{eq:basis.4}) and (\ref{eq:basis.5}) constitute the central result of this manuscript.
They allow us to calculate the polaron wavefunction without resorting to supercell calculations,
but only starting from standard ingredients of DFT calculations in the unit cell, such as KS
states, phonons, and electron-phonon matrix elements.\cite{Baroni2001,Giustino2017}

\subsection{Lattice distortion in the polaronic ground state}\label{sec.pol-eqs-h}

The polaron eigenvector $A_{n\bk}$ obtained from the solution of Eqs.~(\ref{eq:basis.4}), (\ref{eq:basis.5})
can be used to find the atomic displacements in the polaron ground state. To this aim we replace
Eqs.~(\ref{eq:basis.1})-(\ref{eq:basis.3}) and (\ref{eq:basis.5}) inside Eq.~(\ref{eq:DFT.9}). After
some manipulations we obtain:
  \begin{equation}
  \label{eq:basis.6}
  \Delta\tau_{\k \a p} = - \frac{2}{N_p}\sum_{\bq\nu} B^*_{\bq\nu} 
  \left(\frac{\hbar}{2M_\k \w_{\bq\nu}}\right)^{\!\!1/2}\!\!\!\!
  e_{\k\a,\nu}(\bq) \, e^{i\bq\cdot \bR_p}.
  \end{equation}
Here we can see that the quantity $B_{\bq\nu}$ has the physical meaning of the amplitude of the phonon
mode $\bq\nu$ which contributes to the atomic displacement $\Delta \tau_{\k\a p}$.
As in the case of the electron wavefunction in the previous section, 
 it is easy to verify that the atomic displacements $\Delta\tau_{\k \a p}$
are real-valued as a result of time-reversal symmetry, $B^*_{\bq\nu}=B_{-\bq+\bG,\nu}$. 
By inverting Eq.~(\ref{eq:basis.6}) we also find that $B_{\bq\nu}$ fulfils the sum rule:
  \begin{equation}
  \label{eq:basis.6b}
  \frac{1}{N_p} \sum_{\bq\nu} \frac{|B_{\bq\nu}|^2}{\w_{\bq\nu}} = 
      \sum_{\k\a p} \frac{M_\k}{2 \hbar} |\Delta \tau_{\k\a p}|^2,
  \end{equation}
where the r.h.s. can be interpreted as a measure of the lattice distortion.

\subsection{Formation energy in the basis of Kohn-Sham states and phonon modes}\label{sec.form-en}

In analogy with Eq.~(\ref{eq:basis.6}) we can derive the formation energy in terms of the eigenvector
$A_{n\bk}$. To this aim we combine Eq.~(\ref{eq:formation.2}) with
Eqs.~(\ref{eq:basis.1})-(\ref{eq:basis.3}) and (\ref{eq:basis.5}). The result is:
  \begin{equation}
  \label{eq:basis.7}
  \Delta E_f = \!\frac{1}{N_p} \!\sum_{n\bk} 
  |A_{n\bk}|^2 (\ve_{n\bk}-\ve_{\rm CBM}) 
  -\frac{1}{N_p}\!\sum_{\bq\nu}|B_{\bq\nu}|^2\hbar\w_{\bq\nu},
  \end{equation}
or equivalently, using Eq.~(\ref{eq:formation.4}):
  \begin{equation}
  \label{eq:basis.8}
  \Delta E_f
  = \ve-\ve_{\rm CBM} + \frac{1}{N_p}\sum_{\bq\nu}|B_{\bq\nu}|^2\hbar\w_{\bq\nu}.
  \end{equation} 
The formation energy in Eq.~(\ref{eq:basis.7}) is composed of one term associated with the electron part 
of the polaron, described by $A_{n\bk}$, and one term associated with the phonon part, described 
by $B_{\bq\nu}$. 
By comparing Eqs.~(\ref{eq:basis.8}) and (\ref{eq:formation.5}) we see that $|B_{\bq\nu}|^2\hbar\w_{\bq\nu}$
represents the contribution of every vibrational mode to the elastic energy of the polaron.
Therefore it is natural to interpret $|B_{\bq\nu}|^2$ as the number of phonons in each mode participating
to the polaron. This heuristic interpretation can be placed on more rigorous ground by moving from a 
classical to a quantum-mechanical description of the ionic coordinates, and by performing a Bogoliubov
transformation.\cite{Bogolubov2014} For now we limit ourselves to emphasize that in DFT calculations
the nuclei are described in the adiabatic and classical approximation, therefore we do not strictly
have phonon quanta in our formalism. By introducing the spectral functions:
  \begin{eqnarray}
  \label{eq:basis.9}
   A^2(E)  &=& \frac{1}{N_p} \!\sum_{n\bk} |A_{n\bk}|^2 \d(E-\ve_{n\bk}+\ve_{\rm CBM}),  \\
   B^2(E)  &=& \frac{1}{N_p}\sum_{\bq\nu}|B_{\bq\nu}|^2\, \d(E-\hbar\w_{\bq\nu}), \label{eq.basis:10}
  \end{eqnarray}
Eq.~(\ref{eq:basis.7}) is recast as:
  \begin{equation}
  \label{eq:basis.11}
  \Delta E_f = \int_0^\infty \!\! A^2(E)\,E\, dE  
       - \int_0^\infty \! B^2(E)\,E\, dE.
  \end{equation}
From these relations we see that the spectral functions $A^2(E)$ and $B^2(E)$ play a similar role
in the polaron problem as the Eliashberg function in the theory of superconductors.\citep{Allen1983}
In Sec.~\ref{sec.spectral} we will show that these functions can be used 
to identify the EPI mechanisms leading to the formation of polarons.

\subsection{Visualization of the polaron wavefunction}\label{sec.wfc} 

In order to visualize the polaron wavefunction $\psi$ in Eq.~(\ref{eq:basis.1}), it is convenient
to resort to a Wannier function representation. Using the standard notation introduced in 
Ref.~\onlinecite{Marzari2012}, each KS state can be expanded in a basis of maximally-localized 
Wannier functions as follows: 
  \begin{equation}
  \label{eq:wfc.1}
  \psi_{n\bk}(\br) = \frac{1}{\sqrt{N_p}}
  \sum_{mp} e^{i\bk\cdot \bR_{p}} U^\dagger_{mn\bk} \wf_m (\br-\bR_p),
  \end{equation}
where $\wf_m (\br)$ is a Wannier function in the unit cell at the origin of the reference frame,
normalized in the supercell, and $U^\dagger_{mn\bk}$ is the unitary matrix that generates the smooth 
Bloch gauge. By combining Eq.~(\ref{eq:basis.1}) and Eq.~(\ref{eq:wfc.1}) we obtain:
  \begin{equation}
  \label{eq:wfc.2}
  \psi(\br) = \sum_{mp} A_m(\bR_p)\, \wf_m (\br-\bR_p),
  \end{equation} 
having defined:
  \begin{equation}
  \label{eq:wfc.3}
  A_m(\bR_p) = \frac{1}{N_p}\sum_{n\bk} e^{i\bk\cdot \bR_{p}}\, U^\dagger_{mn\bk}\, 
  A_{n\bk}.
  \end{equation}
Equation~(\ref{eq:wfc.2}) naturally defines $A_m(\bR_p)$ as the envelope function of the polaron, starting 
from an {\it ab initio} perspective. It is interesting to observe that Eq.~(\ref{eq:wfc.3}) for the 
electron part of the polaron is entirely analogous to Eq.~(\ref{eq:basis.6}) for the phonon part. 
Equation~(\ref{eq:wfc.3}) is also useful for practical calculations, especially in combination with 
Wannier-Fourier interpolation of the electron-phonon matrix elements, as we will show in 
Sec.~\ref{sec.wavefunction}.

It should be noted that the use of Eq.~(\ref{eq:wfc.2}) requires some care: 
the KS wavefunctions employed to determine $A_{n\bk}$ from Eqs.~(\ref{eq:basis.4}) and 
(\ref{eq:basis.5}) must be the {\it same} as those employed to construct maximally-localized Wannier 
functions, i.e. the matrix $U^\dagger_{mn\bk}$ required in Eq.~(\ref{eq:wfc.3}). Failure to do so would 
result in the introduction of spurious phases and the calculation of an incorrect envelope function.

If the Wannier functions are real, and the wavefunctions $\psi_{n\bk}$ fulfil time-reversal symmetry
($\psi_{n,-\bk} = \psi^*_{n\bk}$, this is not automatically guaranteed in {\it ab initio} calculations), 
then it follows that $U_{mn,-\bk} = U_{mn\bk}^*$. Combined with Eq.~(\ref{eq:wfc.3}), these properties 
imply that also the envelope functions $A_m(\bR_p)$ will be real-valued.

By combining Eqs.~(\ref{eq:wfc.3}) and (\ref{eq:basis.1b}) we obtain the normalization condition on the
envelope function:
  \begin{equation}
  \label{eq:wfc.4}
  \sum_{mp} |A_m(\bR_p)|^2 = 1,
  \end{equation}
where we used the property that $U_{mn\bk}$ is a unitary matrix.

\subsection{Link with the Landau-Pekar model}\label{sec.dft-lp}

We now show that, under suitable approximations, the {\it ab initio} polaron equations 
Eqs.~(\ref{eq:basis.4})-(\ref{eq:basis.5}) reduce precisely to the LP model discussed in
Sec.~\ref{sec.lp}.

To this aim we consider a model system with only one conduction band with effective mass $m^*$, 
one dispersionless phonon mode with frequency $\w_{\text{LO}}$, and electron-phonon coupling given by
the Fr\"ohlich interaction. The electron-phonon matrix element $g(q)$ is given 
by:\citep{Verdi2015,Giustino2017,Giustino2019}
  \begin{equation}
  \label{eq.abinit-to-LP.9}
  |g(q)|^2 = \frac{e^2}{4\pi\e_0}\frac{4\pi}{\Omega}\frac{\hbar\w_{\text{LO}}}{2}\frac{1}{\k\,q^2}.
  \end{equation}
This expression is valid for an isotropic crystal with a single infrared-active phonon.
By replacing Eqs.~(\ref{eq:wfc.3}) and (\ref{eq.abinit-to-LP.9}) inside 
Eqs.~(\ref{eq:basis.4})-(\ref{eq:basis.5}), after some algebra 
we obtain:
  \begin{eqnarray}
  \label{eq.abinit-to-LP.6}
  &&-\frac{\hbar^2\nabla^2}{2m^*} A(\bR) 
  - \sum_{\bR'} \frac{1}{N_p}\sum_{\bq}  e^{i\bq\cdot(\bR'- \bR)} 
  \frac{2}{\hbar\w} |g(q)|^2 |A(\bR')|^2 \nonumber \\ && \hspace{30pt}\times A(\bR) = \ve \,A(\bR),
  \end{eqnarray}
where we omitted the subscript $p$ from $\bR_p$ for notational simplicity, and
the gradient is with respect to $\bR$.

In the limit of dense Brillouin-zone sampling, i.e. supercell of infinite size,
we can replace the summation over $\bq$ by an integral using
$N_p^{-1}\sum_\bq = \Omega_{\rm BZ}^{-1}\int_{\rm BZ} d\bq$. Using this replacement and carrying
out the integral, Eq.~(\ref{eq.abinit-to-LP.6}) becomes:
  \begin{equation}
  \label{eq.abinit-to-LP.11}
   -\frac{\hbar^2 \nabla^2}{2m^*} A(\bR) - \frac{e^2}{4\pi\e_0} \frac{1}{\k} 
   \sum_{\bR'} \frac{\,\,|A(\bR')|^2}{|\bR'-\bR|} A(\bR) = \ve\, A(\bR).
  \end{equation}
We can now transform the summation over the lattice vectors into an integral, 
by regarding $\bR$ as a continuous variable and using the substitution $\Omega \sum_\bR 
= \int d\bR$:
  \begin{equation}
  \label{eq.abinit-to-LP.12}
  -\frac{\hbar^2 \nabla^2}{2m^*} A(\bR)
   - \frac{e^2}{4\pi\e_0} \frac{1}{\k} 
   \frac{1}{\Omega}\int d\bR^\prime \frac{\,|A(\bR^\prime)|^2}{|\bR^\prime-\bR|}
  A(\bR^\prime) = \ve\, A(\bR).
  \end{equation}
By comparing this result with Eq.~(\ref{eq:LP.7}), we see that the envelope function 
$\Omega^{-1/2}A(\br)$ coincides with the solution $\psi(\br)$ of the LP model. Therefore,
in the case of single-band and single-phonon isotropic systems with Fr\"ohlich electron-phonon 
coupling, there exists a direct and unambiguous link between the LP model and first-principles 
calculations of polarons.

\section{Quadratic self-interaction correction for polarons}\label{sec.sic}

As anticipated in Sec.~\ref{sec.abinitio}, the fifth and sixth lines of Eq.~(\ref{eq:DFT.5}) contain
Hartree and exchange-correlation self-interaction energy of the polaron wavefunction. These terms
are a DFT artefact and in a more accurate many-body picture the excess electron should not 
interact with itself. The practical consequence of having these terms is that they prevent electron
self-trapping. For example it is immediate to see that the Hartree term always decreases the formation energy of the polaron. As we show in Sec.~\ref{sec.validation}, we confirmed
by direct calculations that we are unable to obtain stable self-trapped polarons in the presence of
these spurious self-interactions. This behavior is also well documented in the 
literature.\citep{Mauri2005,Sadigh2015}

In order to remove the polaron self-interaction terms in Eq.~(\ref{eq:DFT.5}), we introduce a modified
DFT functional with SIC as follows:
  \begin{eqnarray} 
  \label{eq:SIC.1}
  &&E^{\rm SIC}[n_\up+\Delta n,n_\down] = E[n_\up+\Delta n,n_\down] -E_{\rm H}[\Delta n -\Delta n_{\rm B}] 
  \nonumber \\ && 
  -\frac{1}{2}\big( E_{xc}[n_\up+\Delta n, n_\down] \!-\! 2 E_{xc}[n_\up, n_\down] \!+\! E_{xc}[n_\up -  
  \Delta n, n_\down] \big), \nonumber \\
  \end{eqnarray}
where $E[n_\up+\Delta n,n_\down]$ 
is a standard DFT functional, as in Eq.~(\ref{eq:DFT.5}). The term $E_{\rm H}$ in this equation 
indicates the Hartree energy functional, and $\Delta n_{\rm B} = (N_p\Omega)^{-1}$ is the compensating 
jellium background.  The form of the functional $E^{\rm SIC}$ is chosen in such a way as to cancel
{\it exactly} the Hartree self-interaction of the polaron, and to cancel the exchange-correlation 
self-interaction up to {\it third order} in the polaron density $\Delta n = |\psi|^2$. In fact, upon 
functional differentiation of the exchange-correlation terms in Eq.~(\ref{eq:SIC.1}) we find:
  \begin{eqnarray}
  \label{eq:SIC.2}
   && E^{\rm SIC}[n_\up+\Delta n,n_\down] = E[n_\up+\Delta n,n_\down] \nonumber \\[4pt] && 
    - \frac{1}{2} \frac{e^2}{4\pi \e_0} \sum_\bT \int d\br d\brp \frac{
    [\Delta n(\br)-\Delta n_{\rm B}][\Delta n(\brp)-\Delta n_{\rm B}]}{|\br-\brp-\bT|} \nonumber \\
    && -\frac{1}{2}\int d\br d\brp \frac{\d^2 E_{xc}}{\d n^\up \d n^\up} \Delta n(\br)\Delta n(\brp) 
     + \mathcal{O}(\Delta n^4),
  \end{eqnarray}
which corresponds precisely to the functional in Eq.~(\ref{eq:DFT.5}) with the fifth and sixth lines removed.
The present analysis demonstrates that, not only our starting functional [as defined by the first four lines
of Eq.~(\ref{eq:DFT.5})] is physically motivated, but also it can be derived from a simple
self-interaction-free DFT functional, as given by Eq.~(\ref{eq:SIC.1}). This is particularly
useful for benchmarking our formalism against direct calculations in large supercells. 

In order to generate KS equations starting from Eq.~(\ref{eq:SIC.1}), we evaluate the functional
derivatives with respect to $\psi_{v\bk\up}$, $\psi_{v\bk\down}$, and $\psi$. As a reminder we have 
$n^\up = \sum_{v\bk} |\psi_{v\bk\up}|^2$, $n^\down = \sum_{v\bk} 
|\psi_{v\bk\down}|^2$, and $\Delta n = |\psi|^2$. The total density is $n = n^\up+\Delta n + n^\down$, 
the spin-up density is $n^\up+\Delta n$, and the spin-down density is $n^\down$. We find the following
modified KS Hamiltonians for spin-up valence electrons ($\hH^{\rm SIC}_{v\up}$), spin-down
valence electrons ($\hH^{\rm SIC}_{v\down}$), and the polaron wavefunction ($\hH^{\rm SIC}_{\rm pol}$):
  \begin{eqnarray}
  \label{eq:SIC.3}
   \hH^{\rm SIC}_{v\up} &=& \hH^{\rm KS}_{\up}[n_\up+\Delta n, n_\down] 
    + V_{xc}^\up[n_\up, n_\down]  \nonumber \\ 
   &-&\frac{1}{2}V_{xc}^\up[n_\up+\Delta n, n_\down] 
   - \frac{1}{2}V_{xc}^\up[n_\up - \Delta n, n_\down], \\
  \hH^{\rm SIC}_{v\down} &=& \hH^{\rm KS}_{\down}[n_\up+\Delta n, n_\down] 
    + V_{xc}^\down[n_\up, n_\down]  \nonumber \\
  &-&\frac{1}{2}V_{xc}^\down[n_\up+\Delta n, n_\down] 
   - \frac{1}{2}V_{xc}^\down[n_\up - \Delta n, n_\down], \\
  \hH^{\rm SIC}_{\rm pol} &=&
  \hH^{\rm KS}_\up[n_\up+\Delta n, n_\down] - V_{\rm H}[\Delta n-\Delta n_{\rm B}] \nonumber \\ 
  &-& \frac{1}{2} V_{xc}^\up[n_\up+\Delta n, n_\down] +\frac{1}{2} V_{xc}^\up [n_\up - \Delta n, n_\down],
  \hspace{20pt}
  \end{eqnarray}
where the Hartree potential $V_{\rm H}$ and the exchange-correlation potentials $V_{xc}^{\up,\down}$ 
are defined in the usual way. In order to avoid false minima which are typically encountered in
self-interaction corrected DFT,\cite{Goedecker1997,Mauri2005} we follow the method of 
Ref.~\onlinecite{Mauri2005} and choose to perform a constrained total energy minimization with
the constraint $\psi_{v\bk\up} = \psi_{v\bk\down}$. The added advantage of this choice is that it
is fully consistent with the assumptions that we used in Sec.~\ref{sec.pol-eqs} to derive
the polaron equations.

Our functional $E^{\rm SIC}$ in Eq.~(\ref{eq:SIC.1}) is similar, albeit not identical, to
the SIC proposed in Ref.~\onlinecite{Mauri2005}. In that work
the authors studied the self-trapping of holes in $\a$-quartz, by using a damped Car-Parrinello 
minimization of the total energy. Using the present notation, their functional reads:
  \begin{eqnarray}
  \label{eq:SIC.4} 
  &&E^\text{\rm SIC, Ref.\scriptsize\onlinecite{Mauri2005}\normalsize}[n_\up+ \Delta n,n_\down] = E[n_\up+ \Delta n,n_\down] 
  -E_{\rm H}[\Delta n] \nonumber \\
  && \hspace{40pt}-E_{xc}[n_\up+\Delta n, n_\down] + E_{xc}[n_\up,n_\down].
  \end{eqnarray}
By comparing this expression with Eq.~(\ref{eq:SIC.1}), we see that the Hartree self-interaction
is removed in a similar way in both approaches. 
The difference lies in the exchange-correlation self-interaction: 
by expanding $E_{xc}[n_\up+\Delta n, n_\down]$ in Eq.~(\ref{eq:SIC.4}) using the functional derivative,
we see that the polaron does not experience the exchange-correlation interaction with 
the valence electrons. This can lead to artificially large band gaps.
By applying the SIC to quadratic order in $\Delta n$ both for the Hartree 
and for the exchange-correlation contributions via Eq.~(\ref{eq:SIC.2}), the polaron experiences 
the usual exchange-correlation interaction with the valence electrons, and band gaps remain unaffected.

The correction provided by Eq.~(\ref{eq:SIC.1}) is easy to implement in DFT schemes which 
perform a direct minimization of the energy functional, and requires minimal changes to existing 
codes.\cite{Giannozzi2017}

As we show in Sec.\ref{sec.convergence}, the functional $E^{\rm SIC}$ defined by Eq.~(\ref{eq:SIC.1})
overcomes the delocalization problem of DFT, and correctly yields localized polaron wavefunctions in
polar materials. Importantly, this method does not require the tuning of Hubbard corrections in DFT+$U$ 
or the mixing parameter $\alpha$ in hybrid functional calculations, since the self-interaction error
is removed from the outset without introducing additional parameters.

\section{Implementation and computational setup}\label{sec.setup}

\subsection{Density-functional theory calculations}\label{sec.dft}

In order to demonstrate the theory developed in Sec.~\ref{sec.abinitio} we perform DFT calculations
using planewaves and pseudopotentials, as implemented in the Quantum ESPRESSO materials simulation
suite,\cite{Giannozzi2009} together with the wannier90\cite{Mostofi2014} and EPW\cite{Ponce2016} codes.
The polaron equations described in Sec.~\ref{sec.pol-eqs-ks} are implemented in a modified version of 
the EPW code, and the visualization of the polaron wavefunctions as described in Sec.~\ref{sec.wfc} 
is performed using a modified version of the wannier90 code and VESTA for visualization.\cite{Momma2011} 
We use the generalized gradient approximation to DFT of Perdew, Burke, and Ernzerhof (PBE),\cite{Perdew1996}
 and optimized norm-conserving Vanderbilt
(ONCV) pseudopotentials,\cite{Hamann2013} with planewaves kinetic energy cutoffs of 150~Ry, 105~Ry, and
70~Ry for LiF, Li$_2$O$_2$, and $\alpha$-SiO$_2$, respectively. In the ground-state
calculations we sample the Brillouin zone with $\Gamma$-centered uniform meshes of size 
$12 \times 12 \times 12 $ and $8 \times 8 \times 8 $ for LiF and Li$_2$O$_2$ respectively, while
$\alpha$-SiO$_2$ is sampled at $\Gamma$. Lattice vectors and internal coordinates are optimized using
this setup before proceeding to calculate polarons. Equations~(\ref{eq:basis.4}) and (\ref{eq:basis.5}) 
require the evaluation of KS energies, phonon energies, and electron-phonon matrix elements on dense uniform
grids. To this aim we employ Wannier-Fourier interpolation,\cite{Giustino2017,Marzari2012,Yates2007} 
as implemented in wannier90 and EPW. In order to validate our approach against explicit
supercell calculations, we consider two systems, Al-doped $\alpha$-SiO$_2$ and Li$_2$O$_2$, and
we perform self-interaction corrected Car-Parrinello calculations using the CP 
code\citep{Laasonen1993} of Quantum ESPRESSO.
The SIC scheme implemented in CP was developed in Ref.~\onlinecite{Mauri2005},
and corresponds to the functional in Eq.~(\ref{eq:SIC.4}). To implement the functional in 
Eq.~(\ref{eq:SIC.2}) we made minor modifications to the existing code.

\subsection{Solution of the polaron equations} \label{sec.solvers}

In order to solve Eqs.~(\ref{eq:basis.4})-(\ref{eq:basis.5}) we rewrite Eq.~(\ref{eq:basis.4})
more conveniently as follows: 
 \begin{equation}
 \label{eq:sol.1}
 \sum_{n'\bk'}H_{n\bk,n'\bk'}\,A_{n'\bk'} = \ve \,A_{n\bk},
 \end{equation}
with
 \begin{equation}
 \label{eq:sol.2}
 H_{n\bk,n'\bk'} = \d_{n\bk,n'\bk'}\ve_{n\bk}
    -\frac{2}{N_p}\sum_\nu B^*_{\bk-\bk',\nu}\,g_{nn'\nu}(\bk',\bk-\bk').
 \end{equation}
In this form it is clear that the solution of Eq.~(\ref{eq:sol.1}) can be obtained using standard
numerical eigensolvers. In order to start the procedure we initialize the vector of coefficients
$A_{n\bk}$ using a Gaussian lineshape centered at the band minimum. From this starting guess we proceed to construct the vector
of coefficients $B_{\bq\nu}$ using Eq.~(\ref{eq:basis.5}). At this point we can set up the Hamiltonian
matrix of Eq.~(\ref{eq:sol.2}) and proceed to the solution of the eigenvalue problem in Eq.~(\ref{eq:sol.1}).
The lowest-energy eigenvector $A_{n\bk}$ is used again in Eq.~(\ref{eq:basis.5}) and the whole
procedure is repeated until convergence in the polaron formation energy as given by Eq.~(\ref{eq:basis.8}).
In all calculations we employ an energy convergence threshold of 0.1~meV.

The $\bk$-point grid employed in Eq.~(\ref{eq:sol.1}) defines the equivalent Born-von K\'arman (BvK)
supercell hosting the polaron. For example a $\bk$-point grid $10\times 10 \times 10$ corresponds
to calculating the polaron wavefunction, the corresponding atomic displacements, and the energetics
in an equivalent $10\times 10 \times 10$ supercell. Since we need information on both $A_{n\bk}$
and $B_{\bq\nu}$, we use the same uniform and $\Gamma$-centered grid for $\bk$-points and $\bq$-points.
When $\bk+\bq$ falls outside of the initial grid, we use the periodic gauge and set $A_{n\bk+\bq} = 
A_{n\bk+\bq+\bG}$, with $\bG$ a reciprocal lattice vector that folds $\bk+\bq$ inside the original
grid. This procedure is necessary to guarantee that the solution vector $A_{n\bk}$
fulfils time-reversal symmetry, see discussion after Eq.~(\ref{eq:basis.5}). 

In the case of large polarons dominated by the Fr\"ohlich coupling, the electron-phonon matrix
elements exhibit a singularity at $\bq=0$.\cite{Verdi2017} As a result the solution vectors
$A_{n\bk}$ tend to have significant weight only in the vicinity of the band extrema. This
is the case of the electron polaron in LiF, for example, as discussed in Sec.~\ref{sec.wavefunction}.
In these situations one needs relatively fine $\bk$- and $\bq$-point meshes, but most grid points
do not contribute to the calculations; to reduce computational cost we use fine grids but we restrict
the Hamiltonian $H_{n\bk,n'\bk'}$ to an inner grid of $\bk,\,\bk'$-points near the band edges. 
We then increase the size of the inner grid to check for convergence.

Since in the present formalism we study a localized charge distribution in a supercell,
the solutions of the eigenvalue problem in Eq.~(\ref{eq:sol.1}) contain a spurious interaction
energy between the polaron and its periodic images. 
The same situation is also found
in the study of charged defects in periodic supercells. In order to eliminate this spurious energy
we employ the standard Makov-Payne correction.\cite{Makov1995} To this aim we perform calculations
for increasing size of the equivalent BvK supercell, and then extrapolate the formation energy
and the polaron eigenvalue using the asymptotic trend $L^{-1}$, where $L$ is the linear size of the
equivalent supercell. For example, in the case of the large electron polaron in LiF, we use
$\bk$-point grids up to $33\times33\times33$. In order to cope with such large grids we use
a distributed-memory eigensolver from the ScaLAPACK library.\citep{Choi1996}

One last aspect that requires some care is the gauge arbitrariness of the electron-phonon
matrix elements $g_{mn\nu}(\bk,\bq)$ that one obtains from Wannier-Fourier interpolation.
The arbitrariness relates to the facts that (i) the unitary rotation $U^\dagger_{mn\bk}$ used 
in Eq.~(\ref{eq:wfc.1}) to go from the smooth Bloch basis to the basis of KS states is determined 
from a separate diagonalization at each $\bk$-point; (ii) the analogous rotation required for the atomic 
displacements in Eq.~(\ref{eq:basis.2}), that is the matrix of vibrational eigenvectors 
$e_{\k\a,\nu}(\bq)$, is also obtained by a separate diagonalization at each $\bq$. 
These diagonalizations have two drawbacks: 1) they do not satisfy the time-reversal symmetry
requirements; 2) they may lead to different results on different architectures, and even
on the same architecture but in different runs. This issue is particularly delicate because,
in order to save memory, we recompute the matrix elements $g_{mn\nu}(\bk,\bq)$ at each self-consistent
iteration. Our benchmarks indicate that this issue can lead to (relatively small) numerical noise in the
calculated formation energies, that shows up as small oscillations in plots of $\Delta E_f$ vs.\ $L$.
In order to eliminate these fluctuations we enforce a predetermined choice for the gauge of 
eigenmodes and wavefunctions, in the same spirit as in Sec.~V\,C of Ref.~\onlinecite{Giustino2007}.
First we rotate $U^\dagger_{mn\bk}$ and $e_{\k\a,\nu}(\bq)$ so that the first nonzero component
is real and positive. Then we check for degeneracies in the electron or phonon energies, and 
we break these degeneracies using a fictitious perturbation. To this aim we set up a Hermitian
perturbation $P_{mn\bk}$ that spans the Bloch subspace. We fill this
matrix by using a sequence of small prime numbers as matrix elements. Then, we diagonalize 
$P'_{ij\bk} = \sum_{mn} U_{im\bk} P_{mn\bk} U^\dagger_{nj\bk}$, where the indices $i$, $j$ are
restricted to the degenerate subspaces. By denoting with $V_{ij\bk}$ the unitary matrix that diagonalizes 
$P'_{ij\bk}$, we construct $U'_{jn\bk} = \sum_p V^{\dagger}_{jp\bk}U_{pn\bk}$. Finally we obtain interpolated
KS states and energies from $U'_{mn\bk}$ instead of $U_{mn\bk}$. If the energies are all non-degenerate,
then we are done. If there are still degeneracies we repeat the operation by filling the
perturbation matrix using the next prime numbers in the sequence.
We note that in the subsequent polaron calculation the KS energies remain 
unaffected by this fictitious perturbation, as from
this procedure we only retain the unitary rotation $V_{mn\bk}$; formally this is equivalent
to taking the limit of a vanishingly small perturbation.
We operate similarly for the vibrational eigenmodes. This procedure guarantees
that all KS states and phonon eigenmodes carry a unique gauge across successive iterations
in the same calculation, or across different machines.
We note that the present procedure is simpler and more efficient than the one used in
Ref.~\onlinecite{Giustino2007}, since here we only perform operations on very small matrices and
we do not calculate explicitly the matrix elements of the fictitious perturbation using planewaves, 
unlike in Ref.~\onlinecite{Giustino2007}.
Finally we enforce time-reversal symmetry by making sure that only half of the $\bk$-points are
effectively employed in Eq.~(\ref{eq:sol.1}), using a simple mapping.

\subsection{Test systems} \label{sec.systems}

\subsubsection{Lithium fluoride}\label{sec.data.lif}

The first test system that we consider is a prototypical ionic insulator, lithium fluoride.
LiF crystallizes in a simple rock-salt structure and is known to be a wide-gap insulator.
As the other members of the alkali halides family, LiF hosts color centers with interesting
optoelectronic properties.\cite{Baldacchini2001} In particular, the V$_{\rm K}$ center is
a self-trapped hole polaron which has been studied in a number of investigations.\cite{Karsai2014,
Pederson1988,Mallia2001,williams1990,shluger1993,Gavartin2003,schirmer2006,Ramo2007,Sadigh2015,kokott2018} 
On the other hand, the electron polaron is expected to be a large polaron and has been investigated 
only by means of model Hamiltonians.\citep{Inoue1970} Here we perform calculations 
for both the small hole polaron and the large electron polaron of LiF, and we show that our formalism
correctly describes both limits on the same footing.

Fig.~\ref{fig1}(a) shows a supercell of LiF (the unit cell consists of only two atoms).
Our optimized lattice parameter is $a=4.058$~\AA, in agreement with the experimental value
 $a=4.02$~\AA.\citep{Dressler1987} Our calculated KS band gap is $E_{\rm g} = 8.9$~eV, and underestimates
the experimental optical gap of 14.2~eV as expected.\cite{Olson1976}
We find isotropic electron and hole effective masses of $0.88\,m_e$ and $3.73\,m_e$,
respectively. The electron mass is in good agreement with the reported values 
0.78-1.2$m_e$,\citep{Page1970,Iadonisi1984} but we could not find previous values for the
hole mass.  The calculated relative dielectric constants are $\epsilon^0 = 10.62$ and $\epsilon^\infty =  2.04$,
to be compared with the measured values $\epsilon^0 = 9.04$ and $\epsilon^\infty = 
1.92$.\citep{Andeen1970,Levin1960} The highest computed phonon energy is $\hbar\omega_{\rm max}=77.0$~meV, 
close to the experimental value $\hbar\omega_{\rm max}= 80$~meV,\citep{Dolling1968} 
and the Fr\"ohlich coupling constant for the electrons is $\alpha = 4.92$.

\subsubsection{Lithium peroxide}\label{sec.data.li2o2}

The second test system that we consider is lithium peroxide, Li$_2$O$_2$. This compound crystallizes 
in a layered hexagonal structure, with space group P6$_3$/mmc. The structure can be though of as
consisting of LiO$_2$ layers intercalated by Li planes as seen on Fig.~\ref{fig1}(b). Li$_2$O$_2$
forms in battery cathodes during the operation of lithium-air batteries, and can degrade
the battery performance through its low electrical conductivity.\cite{Ong2012,Feng2013,Viswanathan2011}
It has been proposed that the low conductivity of this compound originates from a strong electron-phonon
coupling, and several studies reported the calculation of small electron polarons using a supercell 
approach.\cite{Kang2012,Lastra2013,Feng2013} In Ref.~\onlinecite{Feng2013} it was shown that
a small electron polaron can form in a $3 \times 3 \times 2$ supercell, without the use of 
Hubbard corrections or hybrid functionals. This finding suggests that Li$_2$O$_2$ supports
strongly bound small polarons. Furthermore Li$_2$O$_2$ is highly anisotropic. These properties
make lithium peroxide an ideal candidate for testing the limits of our approach. 

Figure~\ref{fig1}(b) illustrates a supercell of this compound: in each unit cell we have
four Li and four O atoms, and the optimized lattice parameters are $a=3.153$~\AA\ and 
$c/a=2.433$. Using these parameters, we calculate a band gap of $E_{\rm g}=2.05$~eV, electron effective
masses in- and out-of-plane of $2.19\,m_e$ and $0.42\,m_e$, respectively, and in-/out-of-plane
relative dielectric constants $\epsilon^\infty = 2.73/3.94$ and $\epsilon^0 = 8.36/14.20$.
The highest phonon energy that we calculate is $\hbar\omega_{\rm max}=98.2$~meV, 
and the in-/out-of-plane Fr\"ohlich coupling constants\cite{Verdi2017} are $\alpha = 4.74/1.54$.
Our calculations are in good agreement with previous ones yielding $a=3.17$~\AA,\citep{Chan2011}
$c/a=2.43$,\citep{Chan2011} $E_{\rm g} = $3.6-4.8~eV,\citep{Garcia2011} and
$\hbar\omega_{\rm max}=99.3$~meV.\citep{Lau2015} 

\subsubsection{$\alpha$-Quartz}

Since in Sec.~\ref{sec.sic} we introduced a modified version of the SIC
for polarons of Ref.~\onlinecite{Mauri2005}, it is important to check that our functional
yields results in line with previous work.\cite{Stokbro2001,Mauri2005,Gudmundsdottir2015}
To this aim we repeat previous calculations on Al-doped $\alpha$-quartz, and we compare the
localization of the trapped hole with the existing results.\cite{Stokbro2001,Mauri2005,Gudmundsdottir2015}

Figure~\ref{fig1}(c) illustrates the optimized structure of the primitive unit cell of
$\alpha$-SiO$_2$, in the absence of the Al defect. Our optimized lattice parameters are 
$a=4.913$~\AA\ and $c/a=1.100$, in agreement with the experimental values
$a=4.904$~\AA\ and $c/a=1.100$.\citep{Smith1963}
We model the defect-induced localized hole using a supercell with 72 atoms, with one Si atom
replaced by Al. The lattice parameters of the supercell are not re-optimized after this
substitution. The defective structure is shown in Fig.~\ref{fig1}(d).

\section{Results} \label{sec.results}

\subsection{Validation of the SIC functional} \label{sec.validation}

\begin{figure*}[t]
  \centering
  \includegraphics[width=0.7\textwidth]{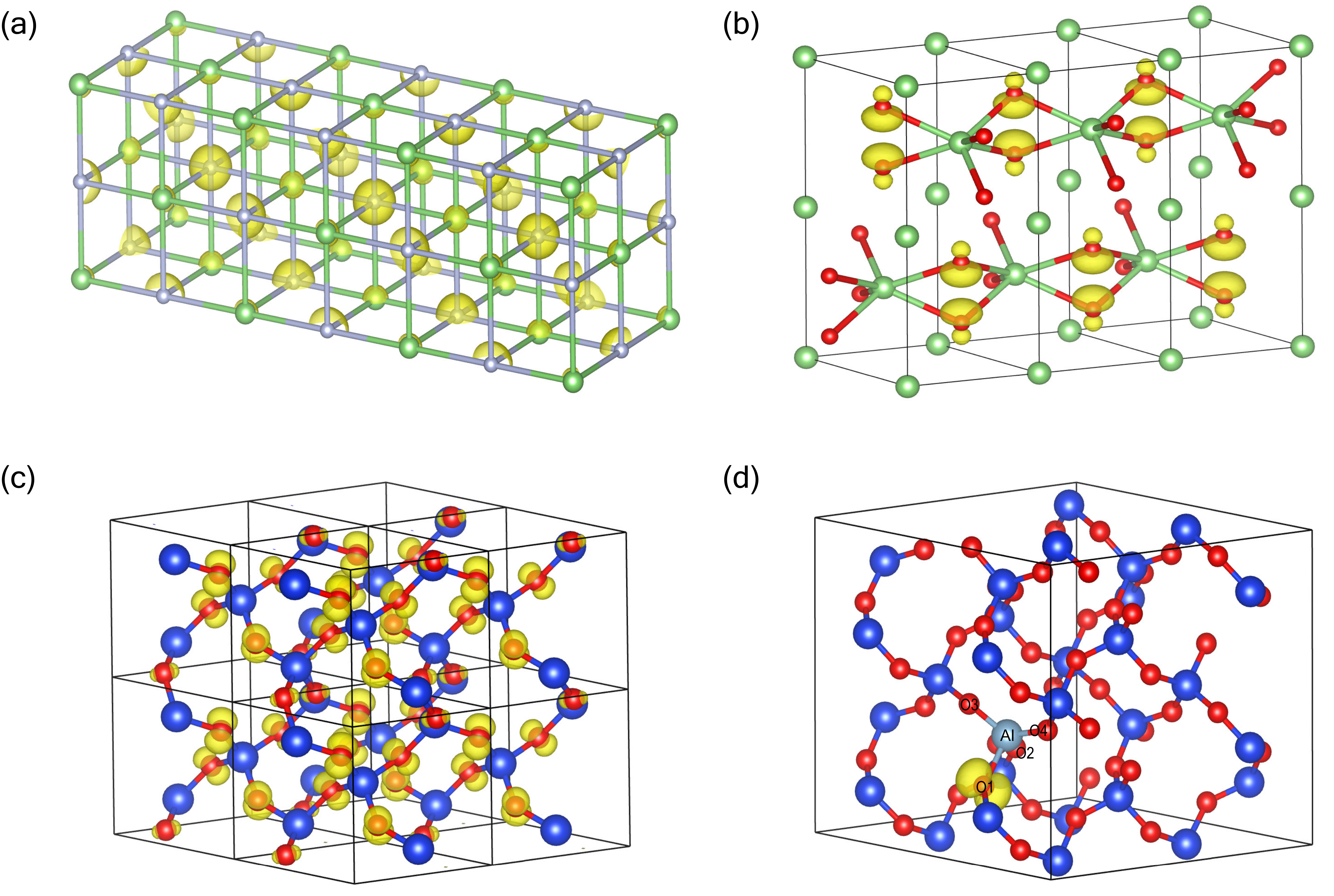}
  \caption{
  Ball and stick models of the compounds considered in this work.
  (a) 3$\times$1$\times$1 supercell of LiF, with Li and F atoms in green and silver, respectively.
  We also show an isosurface plot of the density at the conduction band bottom. In the undistorted
  structure this state is completely delocalized.
  (b) 3$\times$1$\times$1 supercell of Li$_2$O$_2$, with Li and O atoms in green and red, respectively.
  Also in this case we show an isosurface plot of the density at the conduction band bottom,
  in the undistorted structure. The electron is completely delocalized.
  (c) 2$\times$2$\times$2 Supercell of $\alpha$-SiO$_2$, with Si and O atoms in blue and red, respectively. 
  The isosurface plot represents the delocalized conduction band bottom in the undistorted
  structure.
  (d) 2$\times$2$\times$2 supercell of $\alpha$-SiO$_2$ with one Al atom (cyan) replacing Si. 
  In this case the lowest unoccupied state is localized near the defect.
  }
  \label{fig1}
\end{figure*}

In order to validate the SIC functional proposed in Eq.~(\ref{eq:SIC.1}), we consider 
an Al defect in $\alpha$-quartz, following previous work.\cite{Stokbro2001,Mauri2005,Gudmundsdottir2015}
A calculation without SIC yields a delocalized electronic state and no lattice distortion,
as shown in Fig.~\ref{fig1}(c). However, when we include the SIC of Eq.~(\ref{eq:SIC.1}), 
we obtain a localized solution, as seen in Fig.~\ref{fig1}(d). This result is in agreement
with previous work based on the unrestricted Hartree-Fock method\cite{Stokbro2001} and 
other SIC schemes.\cite{Mauri2005,Gudmundsdottir2015}

To be more quantitative we also calculate the bond lengths around the defect site.
Using the labeling convention set out in Fig.~\ref{fig1}(d), our SIC functional yields
the bond lengths 1.946/1.696/1.708/1.699~\AA\ for the bonds Al-O(1) to Al-O(4), respectively.
These values compare well with previous findings, with r.m.s. deviations of only 
0.006~\AA.\cite{Mauri2005,Stokbro2001}
We can conclude that our modified SIC functional yields the same geometry as in previous work.
We also confirmed that the isosurface of the hole density [Fig.~\ref{fig1}(d)] looks
similar to what previously reported.\cite{Stokbro2001,Mauri2005,Gudmundsdottir2015}

To avoid possible ambiguity, we emphasize that the localized hole in Al-doped $\alpha$-SiO$_2$ does not
constitute a polaron strictly speaking. In fact the localization and self-trapping are driven 
by the crystal potential of Al, and do not reflect a spontaneous breaking of translational
symmetry as in the cases of Li$_2$O$_2$ and LiF discussed below. Accordingly, in this case
we do not compare with our linear-response polaron formalism, which addresses spontaneous 
symmetry breaking in perfect crystals.

As a second test we check the geometry of the small electron polaron in Li$_2$O$_2$. In this
case previous work finds an electron localized around two nearest-neighbor O atoms in the LiO$_2$
plane, see for example Fig.~\ref{fig5}(a). The O-O distance in the pristine lattice
is 1.54~\AA~(1.51~\AA~in Ref.~\onlinecite{Kang2012}). Using hybrid functional calculations, 
Ref.~\onlinecite{Kang2012} reported that this distance increases to 2.20~\AA\ upon adding one excess 
electron in a supercell with 192 atoms. Our calculations using the SIC functional
of Eq.~(\ref{eq:SIC.1}) also yield an electron localized around the
same pair of oxygen atoms, as shown in Fig.~\ref{fig5}(e). The resulting O-O distance
is 2.25~\AA, only 2~\% larger than in Ref.~\onlinecite{Kang2012}. 

\subsection{Polaron energy vs.\ supercell size and
Mott transition} \label{sec.convergence}

\begin{figure*}[t]
  \centering
  \includegraphics[width=0.8\textwidth]{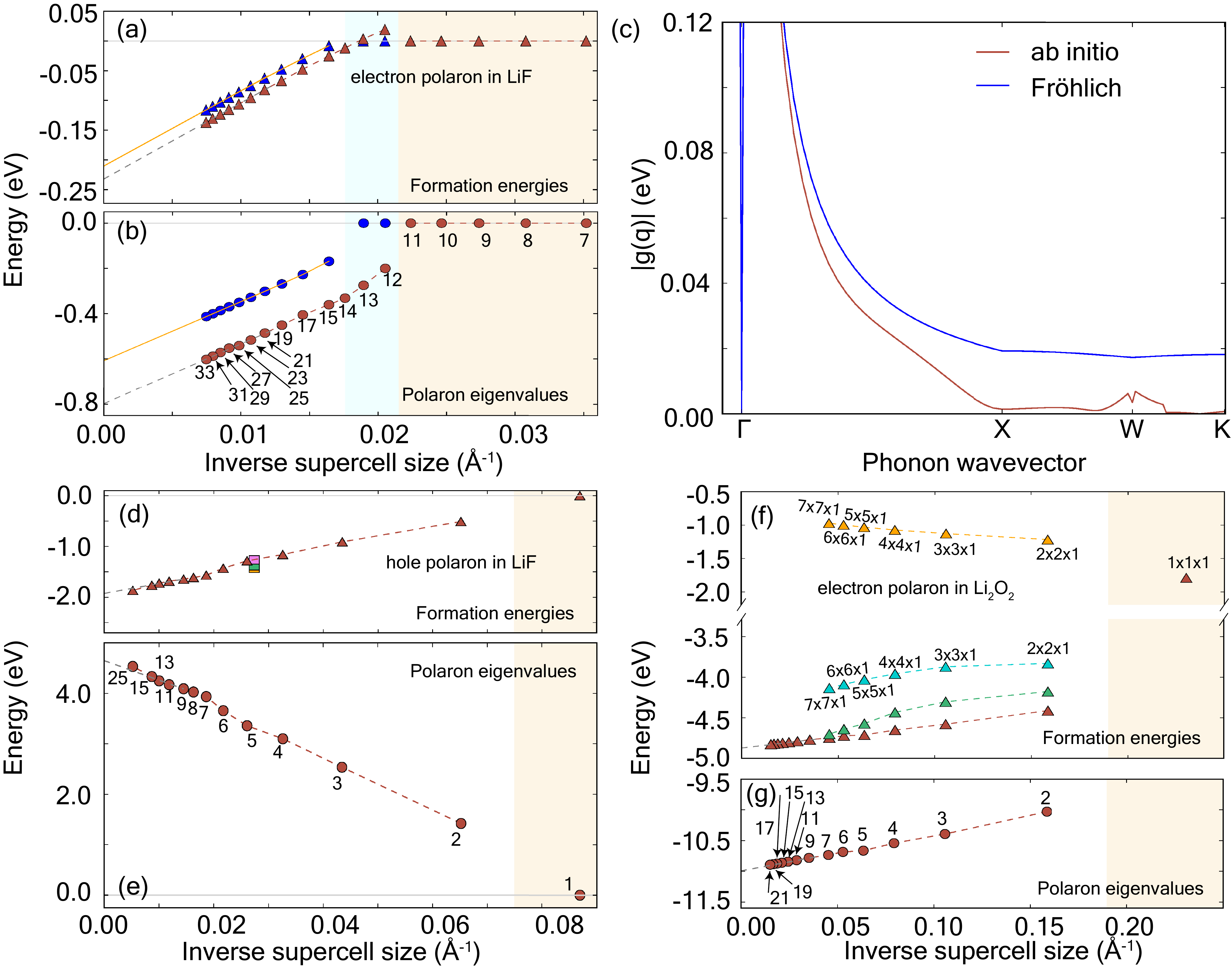}
  \caption{ 
  (a) Formation energy $\Delta E_f$ of the electron polaron in LiF vs.\ supercell size. 
  We give the size as $L^{-1}$, where $L^3$ is 
  the supercell volume. Brown symbols are our calculations using the polaron equations of this
  work. The dashed grey line is the Makov-Payne extrapolation to infinite supercell size.
  The orange line is the result of the LP model. The blue symbols are our calculations after
  considering a parabolic band and a dispersionless LO mode. The shaded regions (blue for parabolic and dispersionless and brown for \textit{ab initio}) indicate supercells
  for which we did not find self-trapped polarons.
  (b) Same as in (a), but this time for the polaron eigenvalue $\ve$. 
  The numbers next to the data 
  points indicate the supercell size, for example 33 means a supercell of size 33$\times$33$\times$33.
  (c) Electron-phonon matrix element for an electron at the conduction band bottom of LiF, as
  a function of the phonon wavevector $|\bq|$. The brown line is the {\it ab initio} matrix element,
  the blue line is the Fr\"ohlich approximation, which retains only the long-range component.
  (d) Formation energy of the hole polaron in LiF vs.\ supercell size (brown symbols). 
  The dashed line is the Makov-Payne extrapolation.
  The filled squares are the formation energies calculated in Ref.~\onlinecite{Sadigh2015}. 
  (e) Same as in (d), but for the eigenvalue of the hole polaron in LiF (brown symbols). 
  The numbers represent the supercell size as in (b).
  (f) Formation energy of the electron polaron in Li$_2$O$_2$ vs.\ supercell size.
  $N\times N\times 1$ indicates a non-uniform supercell (used for computational convenience).
  We compare the results of our polaron equations (green symbols) and our explicit DFT calculations 
  with the SIC functional of Eq.~(\ref{eq:SIC.1}) (cyan symbols), both on non-uniform $N\times N \times 1$
  supercells. We also include DFT calculations without self-interaction correction (orange symbols), and our polaron equations on uniform supercells (brown symbols).
  (g) The eigenvalue of the electron polaron in Li$_2$O$_2$ vs.\ supercell size (brown symbols).
  The notation is the same as in (b).
  }
  \label{fig2}
\end{figure*}

In Fig.~\ref{fig2}(a),(b) we compare the formation energy and polaron eigenvalue obtained
via our Eqs.~(\ref{eq:basis.4})-(\ref{eq:basis.5}) (brown symbols) with the results of the 
continuous LP model described in Sec.~\ref{sec.lp} (orange lines). We focus on the large 
electron polaron in LiF for definiteness, and for calculations using the LP model we 
take $\k = 2.53$ and $m^*/m_e = 0.88$ from Sec.~\ref{sec.data.lif}. Fig.~\ref{fig2}(a) shows 
that the polaron formation energy scales with the supercell size as $L^{-1}$, as expected. 
In the LP model, the formation energy extrapolated at infinite supercell size is 
{$\Delta E_f = -210$}~meV. In contrast, when we solve the {\it ab initio} polaron equations, 
we find the extrapolated energy $\Delta E_f = -231$~meV. The difference between the LP model 
and our method relates to the fact that in our {\it ab initio} calculations the bands, phonons, 
and electron-phonon matrix elements are not as simple as in the LP model. To demonstrate this 
point, we show in the same figure a calculation carried out using our method, but after replacing 
the band structure by a parabolic band with the same effective mass as in the LP mode, the 
phonon dispersion relations by a single, non-dispersive longitudinal-optical (LO) mode, 
and retaining only the long-range component of the electron-phonon matrix element. 
This ``trimmed'' version of the calculation reproduces the LP model exactly,
as shown by the blue symbols in Fig.~\ref{fig2}(a). 
A comparison of the {\it ab initio} electron-phonon matrix element for this mode and the
long-range Fr\"ohlich component used in the LP model is shown in Fig.~\ref{fig2}(c); here we
see that the LP model overestimates the strength of the coupling at short range. 
Besides validating our method, the present comparison highlights the fact that even in a compound 
as simple as LiF the electron-phonon coupling is more complex than a simple Fr\"ohlich interaction,
and that details of band structures, phonon dispersions, and matrix elements are to be taken
into account for predictive calculations. 

In Fig.~\ref{fig2}(b) we report the polaron eigenvalue $\ve$ measured from the conduction-band 
bottom as a function of supercell 
size $L$. In this case the Makov-Payne extrapolation to infinite supercell size yields 
$\ve = -800$~meV with our method (brown symbols), and $\ve = -609$~meV with the LP model 
(orange line). As for the formation energy, also in the case of the polaron eigenvalue we fully
recover the LP result when we consider a parabolic band and a non-dispersive LO phonon 
[blue symbols in Fig.~\ref{fig2}(b)]. It is interesting to note that in LiF the ratio between 
the polaron eigenvalue and its formation energy is 3.46; this ratio 
is close to the prediction of the LP model in Sec.~\ref{sec.lp}, which yields $\varepsilon/\Delta E_f = 3$ 
using the exponential ansatz in Eqs.~(\ref{eq:LP.13}) and (\ref{eq:LP.13b}); note that in the 
LP model $\Delta E_f$ coincides with the energy $E_{\rm LP}$.

In Fig.~\ref{fig2}(a) we also see that when the LiF supercell is smaller than $12\times12\times12$
unit cells there is no localized polaron solution, i.e.\ $\Delta E_f = 0$. The existence of a critical 
supercell size for polaron formation can be explained in terms of the Mott transition: when the periodic 
replicas of the polaron are too close, they form an extended wavefunction, and the corresponding lattice 
deformation is too shallow to trap an electron. In this case the excess electron becomes fully delocalized. 
Therefore a localized polaron can only form when the overlap between nearest-neighbor replicas is negligible. 
This is the same criterion used by Mott to identify the metal-insulator transition.\citep{Mott1968}
Using the Mott criterion in the standard form $r_{\rm p}\, n_{\rm c}^{1/3} = 0.26$,\citep{Mott1968}
with $n_{\rm c}$ being the critical density and $r_{\rm p}$ from Eq.~(\ref{eq:LP.12}), we can estimate 
a critical density:
  \begin{equation}\label{eq.mott}
     n_{\rm c} \simeq 3.6 \left(\frac{m^*/m_e}{\k}\right)^{\!\!3} \cdot 10^{21} \mbox{cm}^{-3}.
  \end{equation}
We note that this is only a {\it crude} estimate since it is based on a simplified solution to the Pekar
polaron problem. Using $\k = 2.53$ and $m^*/m_e = 0.88$ from 
Sec.~\ref{sec.data.lif} inside Eq.~(\ref{eq.mott}), we obtain $n_{\rm c} =15\cdot  10^{19}$~cm$^{-3}$.
This estimate is of the same order of magnitude as our calculation in Fig.~\ref{fig2}(a), 
which places the transition between supercells of size $11^3$ and $12^3$, that is
$n_{\rm c} =4\cdot  10^{19}$~cm$^{-3}$.

In Figs.~\ref{fig2}(d),(e) we show our calculated formation energy and eigenvalue for the
hole polaron in LiF, respectively. In this case we obtain self-trapped polarons already for supercells
as small as $2\times2\times2$ unit cells. This result is consistent with Eq.~(\ref{eq.mott})
and the heavy effective mass of the valence bands. In fact if we use $m^*/m_e = 3.73$ 
from Sec.~\ref{sec.data.lif} we obtain $n_{\rm c} = 1.15\cdot  10^{22}$~cm$^{-3}$,
which corresponds approximately to one electron in a $2\times 2\times 2$ supercell.
The Makov-Payne extrapolation yields $\Delta E_f = -1.98$~eV and $\ve = +4.76$~eV (measured 
from the valence-band top), therefore
we are in the presence of a strongly bound polaron. We note that the polaron eigenvalue is
positive because the localized hole state lies above the valence band, but the energy is still
within the KS gap of this system ($E_{\rm g}= 8.9$~eV from Sec.~\ref{sec.data.lif}).
For comparison with explicit DFT calculations, in Fig.~\ref{fig2}(d) we also report
the formation energies calculated in Ref.~\onlinecite{Sadigh2015} using SIC or hybrid functionals
(filled squares). These calculations correspond to $5\times 5\times5$ supercells and are in
very good agreement with our results.

In Figs.~\ref{fig2}(f),(g) we show the eigenvalues and formation energies of the electron
polaron in Li$_2$O$_2$, respectively, as a function of supercell size. 
Using $\k = 4.05$ and $m^*/m_e = 2.19$ from Sec.~\ref{sec.data.li2o2}
inside Eq.~(\ref{eq.mott}), we obtain the estimate $n_{\rm c} =6\cdot  10^{20}$~cm$^{-3}$; therefore we
expect to see localized solutions already for supercells as small as $5\times 5\times 1$
unit cells. Our calculations indeed find polarons already at $2\times2\times 1$, see Fig.~\ref{fig2}(f).
In this case the formation energy and eigenvalue extrapolated at infinite supercell size (brown symbols) are
$\Delta E_f = -4.87$~eV and $\ve = -10.98$~eV, respectively. The polaron eigenvalue
falls within a band gap in the valence manifold. 
In this figure we also compare to direct calculations using the SIC functional of Sec.~\ref{sec.sic}.
The formation energy in our explicit DFT SIC calculation (cyan symbols) is close to the results of our
linear response polaron equations (green symbols), and exhibits the same trend as a function
of supercell size; the DFT SIC calculation for the largest supercell considered here ($7\times 7\times 1$)
yields $-4.13$~eV, to be compared to our linear-response result $-4.70$~eV. 
The deviation of $\sim$14\%  can be attributed to the fact that our formalism neglects
the response of the valence electrons to the localized lattice distortion caused by this strongly
bound polaron, or to the fact that the approximation of linear electron-phonon coupling becomes
inaccurate for such large atomic displacements. In the same figure we also show that a DFT 
calculation without SIC fails to predict the correct formation energy, and tends to delocalize 
the polaron when increasing the supercell size (orange symbols). We note that a study of the
scaling of the polaron energy vs.\ supercell size has not yet been reported in the literature. 

\subsection{Polaron wavefunctions} \label{sec.wavefunction}

\begin{figure*}[t]
  \centering
  \includegraphics[width=0.9\textwidth]{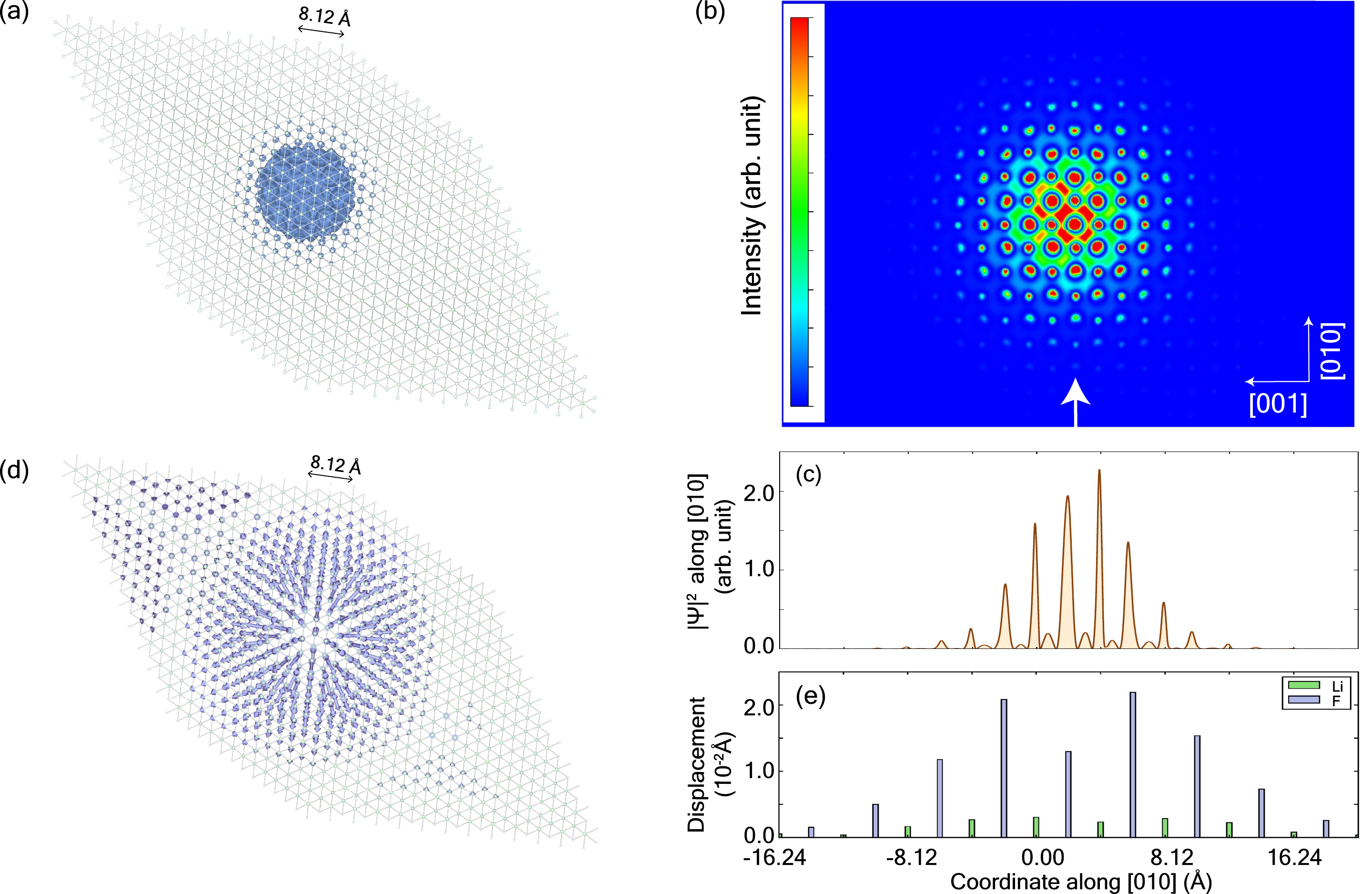}
  \caption{ 
  Electron polaron in LiF.
  (a) Isosurface plot of the polaron wavefunction $\psi$ computed with our method for an extra
  electron in LiF. We use a 12$\times$12$\times$12 supercell, as it can be seen from the
  underlying ball-stick model (Li and F are in green and silver, respectively).
  (b) Same wavefunction as in (a), but as a contour plot in a plane passing through the
  center and perpendicular to the [100] direction.
  (c) One-dimensional profile of the polaron density $|\psi|^2$ along the line indicated by 
  the arrow in (b). 
  (d) Displacements of F atoms in this polaron state. The length of the arrows has been 
  scaled $\times 150$ for visualization purposes. 
  (e) Absolute values of the Li and F displacements along a line passing near the
  polaron center. 
  }
  \label{fig3}
\end{figure*}

Figure~\ref{fig3} shows the electron polaron in LiF, obtained by solving Eqs.~(\ref{eq:basis.4}) and 
(\ref{eq:basis.5}). The electron wavefunction is computed using Eq.~(\ref{eq:wfc.1}), and the atomic
displacements are obtained via Eq.~(\ref{eq:basis.6}). Figures~\ref{fig3}(a),(b) show the electron 
wavefunction as an isosurface and as a contour plot in a plane cutting through the center, respectively. 
When we compare with the delocalized electronic state shown in Fig.~\ref{fig1}(a) we see that now
we are in the presence of a localized, but large, polaron. To quantify the spatial extension of the polaron we
plot the electron density along a line going through the polaron center, see Fig.~\ref{fig3}(c).
The envelope of the resulting function resembles a Gaussian; if we define the polaron size as the
full width at half maximum we obtain $2r_{\rm p} = 9.0$~\AA. Therefore this polaron
extends approximately over two unit cells of the LiF lattice. In Figs.~\ref{fig3}(d),(e)
we report the atomic displacements associated with this polaron, using a vectorial representation
and a one-dimensional cut, respectively. We note that the displacements
of the F anions are consistently larger than those of the Li cations. This may appear counterintuitive because
the F anions are heavier, but it is consistent with the fact that the electron charge is mostly
concentrated around the Li cations due to the character of the 
conduction band bottom, therefore the F atoms experience the strongest electrostatic force. 
The largest atomic displacement is 0.02~\AA, and this value is only 
1\% of the Li-F bond length. Therefore we are well within the remit of the harmonic approximation.

\begin{figure*}[t]
  \centering
  \includegraphics[width=0.9\textwidth]{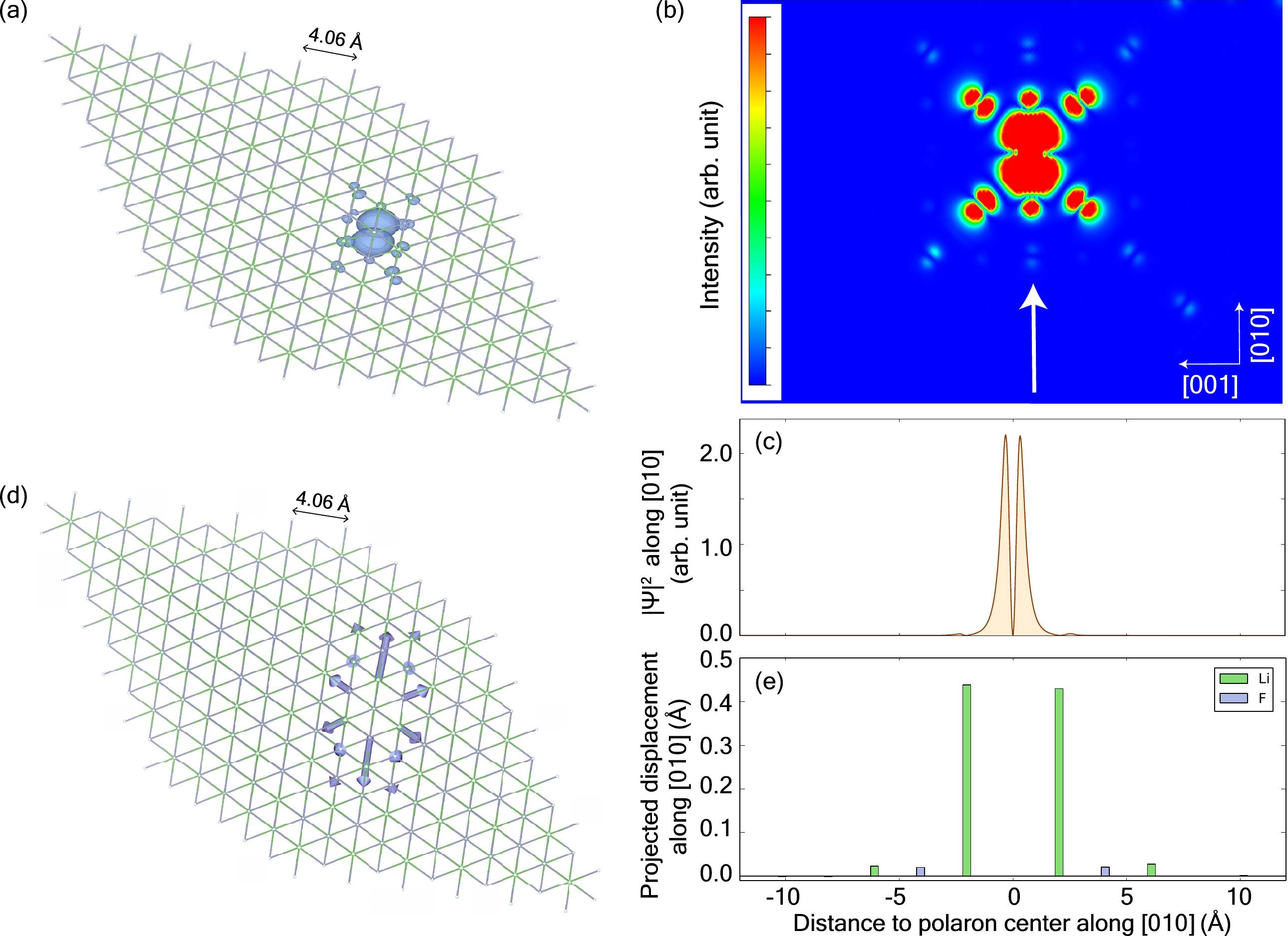}
  \caption{
  Hole polaron in LiF.
  (a) Isosurface plot of the polaron wavefunction $\psi$ computed with our method for an extra
  electron in LiF. We use a 5$\times$5$\times$5 supercell, as shown by the
  underlying ball-stick model (Li and F are in green and silver, respectively).
  (b) Same wavefunction as in (a), but as a contour plot in a plane passing through the
  center and perpendicular to the [100] direction.
  (c) One-dimensional profile of the polaron density $|\psi|^2$ along the line indicated by 
  the arrow in (b). 
  (d) Displacements of Li atoms in this polaron state. The length of the arrows has been 
  scaled $\times 8$ for clarity.
  (e) Absolute values of the Li and F displacements along the same line used in (c).  
  }
  \label{fig4}
\end{figure*}

Figure~\ref{fig4} shows the hole polaron in LiF, namely (a) the wavefunction isosurface, (b) the
same function as a contour plot, (c) a line-cut of the wavefunction, (d) the atomic displacements
as arrows, and (e) the size of the displacements along a line passing through the center.
Here we are in the presence of a small hole polaron, which is expected given the much heavier
masses of the holes as compared to electrons in this system and the much narrowed valence
band width [Fig.~\ref{fig6}(a)]. As it will be discussed in Sec.~\ref{sec.spectral} the hole
polaron in LiF is much closer to a Holstein polaron than a Fr\"ohlich polaron. In this case the 
wavefunction extends over approximately two atomic orbitals, and from Fig.~\ref{fig4}(c) 
we obtain the full width at half maximum $2r_{\rm p} = 0.97$~\AA. Accordingly only a few atoms 
undergo significant displacements, as shown in Fig.~\ref{fig4}(e). The largest displacements are 
found for Li cations, in line with the fact that the wavefunctions at the top of the valence band are 
localized around the anions, which therefore experience a weaker force. 
We obtain a maximum displacement of 0.44~\AA, which is approximately 20\% of the bond length 
of Li-F (2.03~\AA). It is remarkable that our formalism is able to capture this limit of very 
small polaron, even when the atomic displacements are definitely beyond the harmonic regime. We believe 
that the reason why the formalism works is this extreme case is that the distortion caused 
by the small polaron affects only a small portion of the crystal; therefore the use of bands, 
phonons, and electron-phonon matrix elements calculated for the undistorted unit cell does not 
lead to significant inaccuracies.

\begin{figure*}[t]
  \centering
  \includegraphics[width=0.9\textwidth]{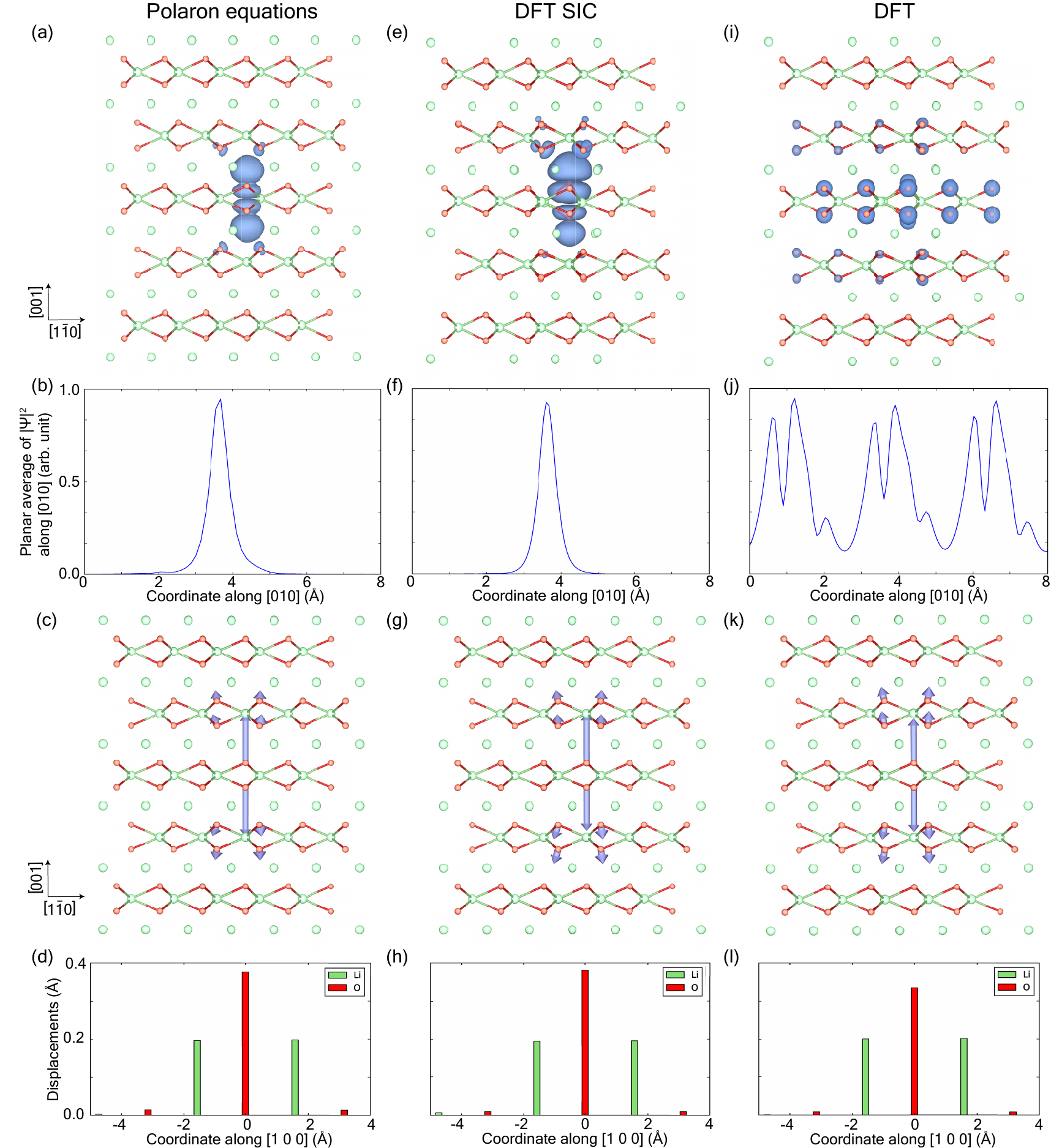}
  \caption{  Electron polaron in Li$_2$O$_2$. All calculations were performed in a 3$\times$3$\times$3 supercell.
  (a) Isosurface plot of the polaron wavefunction $\psi$, computed using our method. The
      green and red spheres are Li and O atoms, respectively.
  (b) Planar average of $|\psi|^2$ along a [010] line passing through the polaron center.
  (c) Displacements of the O atoms in this polaron, amplified $\times$8 for clarity.
  (d) Absolute value of the displacements along a [100] line passing near the polaron center.
  In (e)-(h) we repropose the same set of data, this time using the DFT SIC functional of 
  Eq.~(\ref{eq:SIC.1}). In (i)-(l) we repropose the same set of data, this time using standard
  DFT calculations without SIC.
  }
  \label{fig5}
\end{figure*}

Figure~\ref{fig5} shows the electron polaron in Li$_2$O$_2$. In this case we compare three
calculations: in (a)-(d) we show the small electron polaron obtained with our formalism;
in (e)-(h) we show an explicit supercell calculation using the SIC functional of Sec.~\ref{sec.sic};
in (i)-(l) we show the results of a standard DFT calculation without SIC. In each column we
report, from top to bottom: the electron wavefunction, its one-dimensional cut across the
polaron center, the atomic displacements as arrows, and the one-dimensional cut of these displacements.
The first observation to be made is that standard DFT yields a two-dimensional electronic state
that is localized along the $c$ axis [Fig.~\ref{fig5}(i)] but delocalized in the $ab$ plane.
The SIC leads to electron localization also in the plane, and this is observed both in the
explicit supercell calculation [Fig.~\ref{fig5}(e)] and using our method [Fig.~\ref{fig5}(a)].
The explicit supercell calculation yields a slightly asymmetric wavefunction, while our method
gives a perfectly symmetric polaron. This is an artifact of the constraint 
$\psi_{v\bk\up} = \psi_{v\bk\down}$ used in the SIC calculation, in fact previous work using
hybrid functionals and supercells also found a symmetric polaron,\cite{Kang2012} as in our method.
In this case the polaron is also very small, and extends over two adjacent O-$p$ orbitals.
From Fig.~\ref{fig5}(b) we determine $2r_{\rm p} = 0.63$~\AA, and from Fig.~\ref{fig5}(d) we
find the largest displacement to be 0.38~\AA. Also in this case the atomic displacements
are large ($\sim$25\% of the O-O distance, 1.51~\AA), but our method correctly predicts
the distorted structure as the explicit DFT SIC calculation. This success is remarkable if we 
consider that our theory is based on small displacements and linear electron-phonon interactions.

\subsection{Spectral decomposition of the polaron} \label{sec.spectral}

\begin{figure*}[t]
  \centering
  \includegraphics[width=0.7\textwidth]{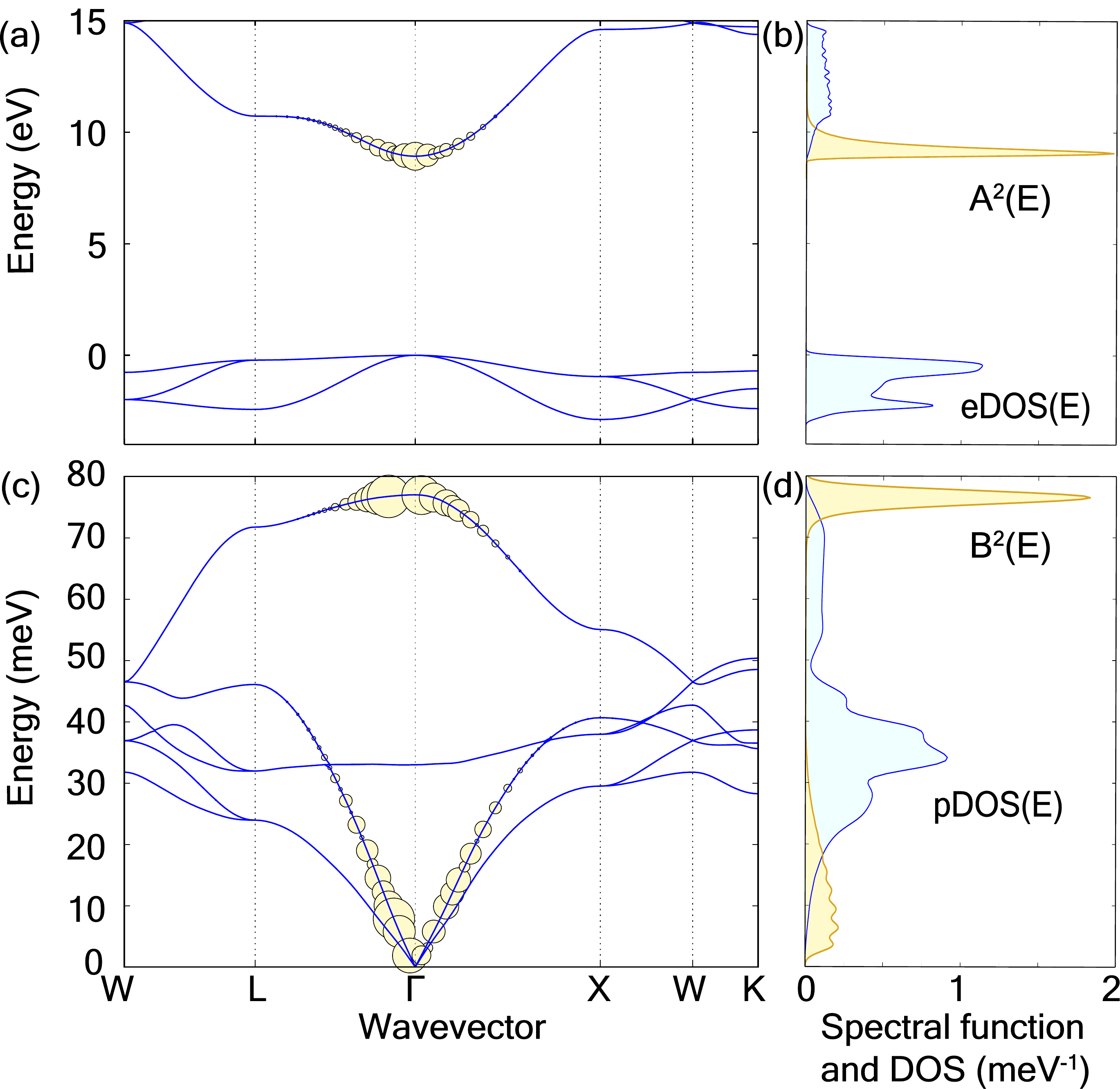}
  \caption{  Spectral decomposition of the electron polaron in LiF.
  (a) Generalized Fourier amplitudes $A_{n\bk}$ plotted on top of the band structure of LiF.
  The radius of each circle is proportional to $|A_{n\bk}|^2$. The zero of the energy is 
  aligned with the top of the valence bands.
  (b) Electronic density of states (blue,~arb. unit) and spectral function $A^2(E)$ (yellow), 
  aligned with the bands in (a).
  (c) Generalized Fourier amplitudes $B_{\bq \nu}$ plotted on top of the phonon dispersion
  relations of LiF. The radius of each circle is proportional to $|B_{\bq\nu}|^2$.
  (d) Phonon density of states (blue,~arb. unit) and spectral function $B^2(E)$, aligned with the
  dispersions in (c).
  }
  \label{fig6}
\end{figure*}

\begin{figure*}[t]
  \centering
  \includegraphics[width=0.7\textwidth]{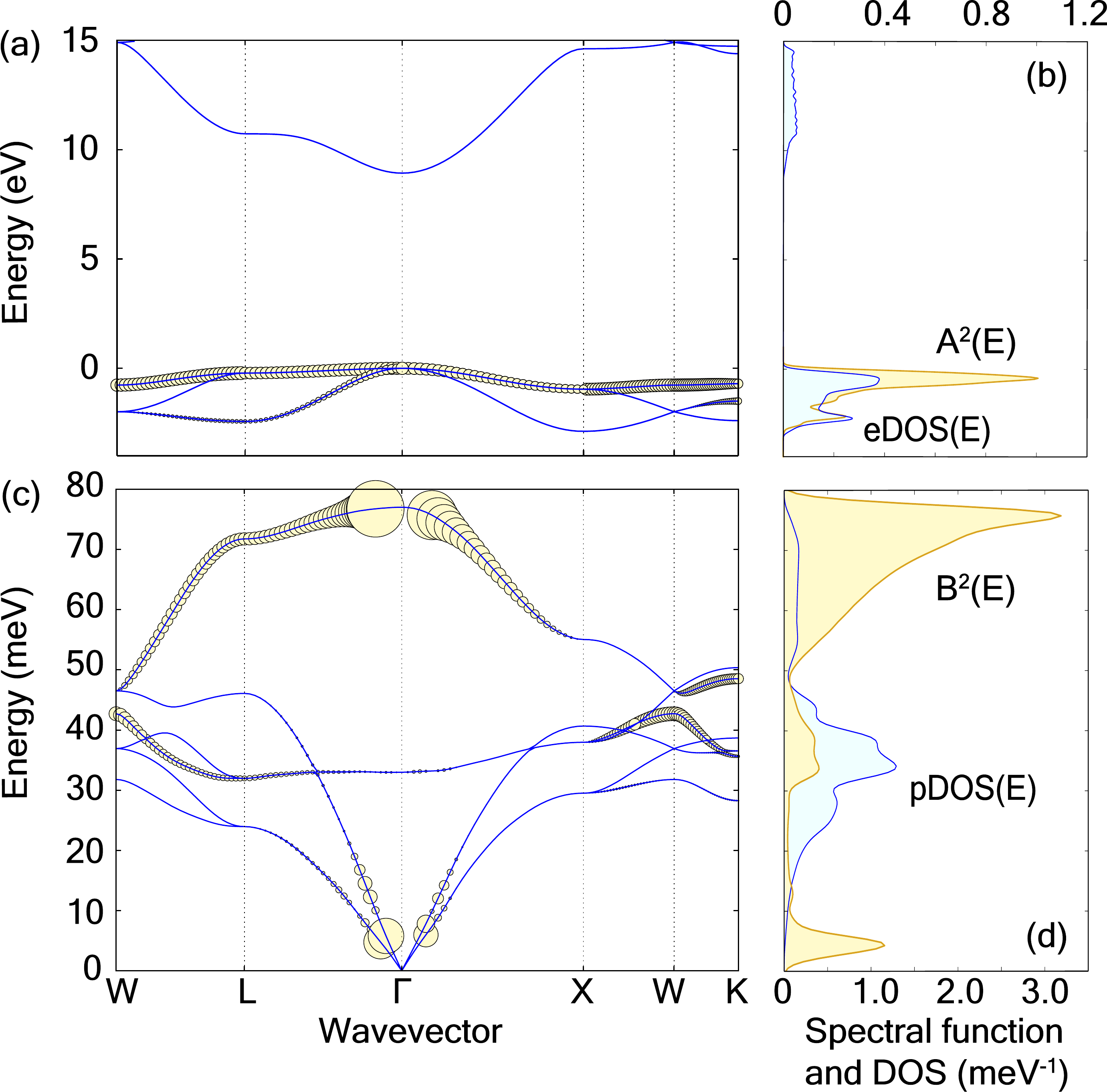}
  \caption{Spectral decomposition of the hole polaron in LiF.
  (a) Generalized Fourier amplitudes $A_{n\bk}$ plotted on top of the band structure of LiF.
  The radius of each circle is proportional to $|A_{n\bk}|^2$.
  (b) Electronic density of states (blue,~arb. unit) and spectral function $A^2(E)$ (yellow), 
  aligned with the bands in (a).
  (c) Generalized Fourier amplitudes $B_{\bq \nu}$ plotted on top of the phonon dispersion
  relations of LiF. The radius of each circle is proportional to $|B_{\bq\nu}|^2$.
  (d) Phonon density of states (blue,~arb. unit) and spectral function $B^2(E)$, aligned with the
  dispersions in (c).
  }
  \label{fig7}
\end{figure*}

\begin{figure*}[t]
  \centering
  \includegraphics[width=0.7\textwidth]{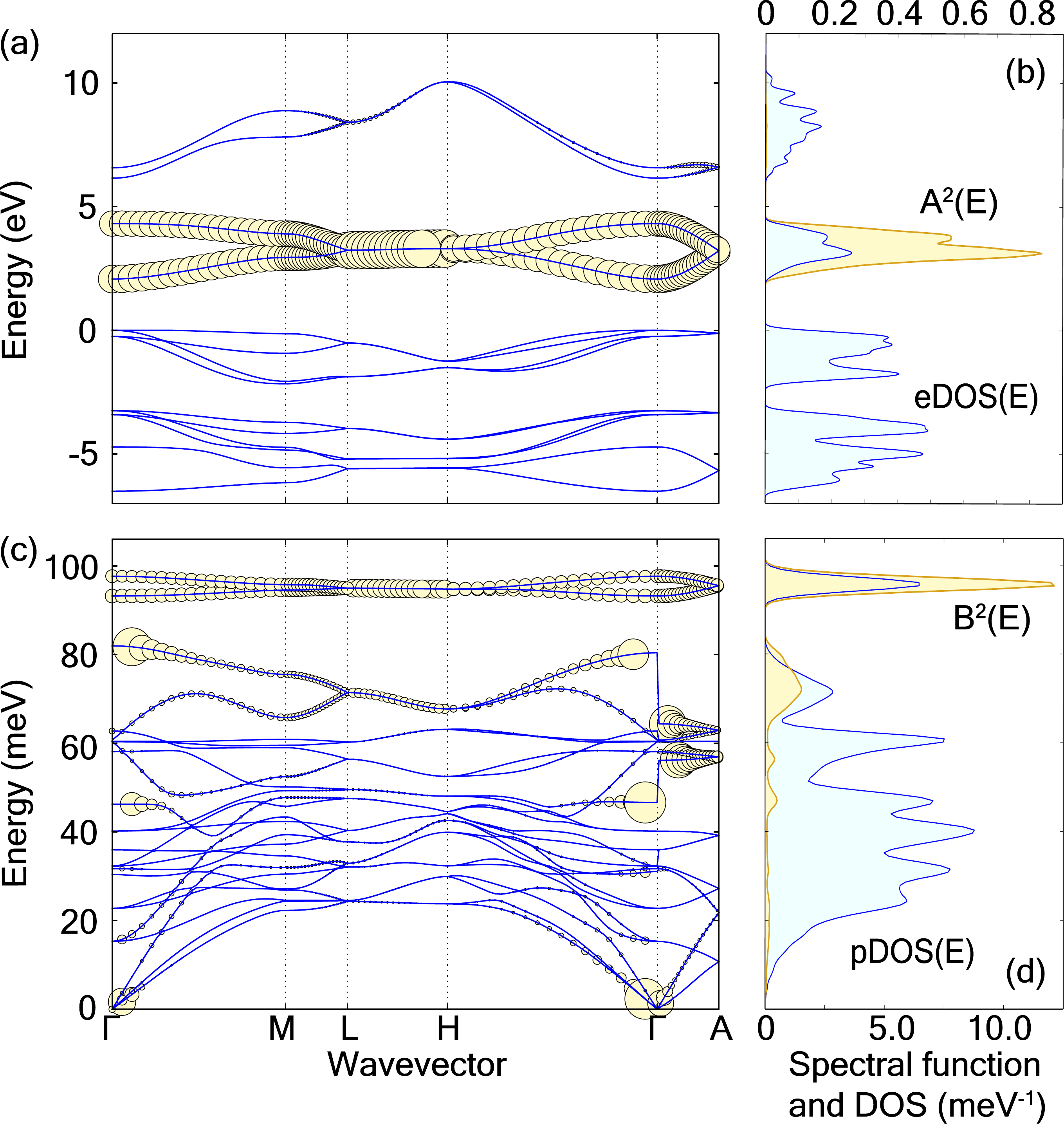}
  \caption{ Spectral decomposition of the electron polaron in Li$_2$O$_2$.
  (a) Generalized Fourier amplitudes $A_{n\bk}$ plotted on top of the band structure of Li$_2$O$_2$.
  The radius of each circle is proportional to $|A_{n\bk}|^2$.
  In this case we scale all radii by a large factor in order to show the tiny contribution
  arising from the topmost unoccupied bands.
  (b) Electronic density of states (blue, arb. unit) and spectral function $A^2(E)$ (yellow), 
  aligned with the bands in (a).
  (c) Generalized Fourier amplitudes $B_{\bq \nu}$ plotted on top of the phonon dispersion
  relations of Li$_2$O$_2$. The radius of each circle is proportional to $|B_{\bq\nu}|^2$.
  (d) Phonon density of states (blue, arb. unit) and spectral function $B^2(E)$, aligned with the
  dispersions in (c).
  }
  \label{fig8}
\end{figure*}

In Figs.~\ref{fig6}-\ref{fig8} we present the spectral decomposition of the polaron wavefunctions
and atomic displacements in terms of the underling band states. Figure~\ref{fig6}(a) shows the
electronic weights $|A_{n\bk}|^2$ plotted on top of the band structure for the case of the
large electron polaron in LiF. The corresponding electronic density of states and spectral
function $A^2(E)$ are shown in Fig.~\ref{fig6}(b). These plots are meant to mimic similar
representations of the excitons calculated via the Bethe-Salpeter method.\cite{Rohlfing2000,Bokdam2016,
Andreoni1995}
The large electron polaron is dominated by states at the bottom of the conduction band; 
as expected the localization in reciprocal space mirrors the delocalization in real space.
The corresponding atomic displacements are resolved using the weights $|B_{\bq\nu}|^2$ in 
Fig.~\ref{fig6}(c), and the density of vibrational states and phonon spectral
function $B^2(E)$ are given in Fig.~\ref{fig6}(d). We see that the polaron is dominated by 
the LO mode at 76~meV, as expected from earlier work on Fr\"ohlich polarons
in halide salts,\citep{Inoue1970} but we also have smaller contributions coming from the acoustic branches.
By integrating $B^2(E)$ in Fig.~\ref{fig6}(d) we can quantify the roles of these phonons:
we find that the LO mode accounts for 62\% of the polaronic distortion, while the
the transverse acoustic (TA) mode is responsible for the remaining 38\%. To the best of
our knowledge this is the first study where the importance of TA phonons in the polaron 
physics of LiF has been identified. This is precisely the kind of new insight that our
method can offer.

Figure~\ref{fig7} shows the spectral decomposition of the small hole polaron in LiF.
Here the main observation is that the entire highest valence band contributes to the
polaronic wavefunction, with smaller contributions from lower-lying bands. This behavior
suggests that the small hole polaron of LiF is closer to the Holstein limit\citep{Holstein1959}
than the Fr\"ohlich limit.\citep{Frohlich1950} We emphasize that, at variance with model
Hamiltonians, our approach is parameter-free, therefore it captures seamlessly both limits. 
Also in this case the LO phonon branch dominates the coupling, however now it is the entire
branch that contributes, as shown in Fig.~\ref{fig7}(d). This observation is in line with
the fact that the small polaron requires short-range electron-phonon coupling, therefore
the range of important phonon wavevectors must extend away from the zone center.
By integrating the spectral function $B^2(E)$ we find that the LO branch contributes
78\% of the coupling in this case.

Finally Fig.~\ref{fig8} shows the spectral decomposition for the small electron polaron
in Li$_2$O$_2$. In this case the two lowest conduction bands contribute equally to the
polaron wavefunctions. This is a case where one-band model Hamiltonians such as the
models of Fr\"ohlich and Feynman would not be sufficient to capture the essential
features of the problem. We also point out that the higher-lying conduction
bands do not contribute appreciably to the polaron wavefunction, as it can be seen 
from the spectral density $A^2(E)$ in Fig.~\ref{fig8}(b). The largest contribution to the 
lattice distortion comes from TO modes around 96~meV, which account for 64\% of the coupling. 

\section{Future developments}\label{sec.future}

Having established the potential of our new methodology in Sec.~\ref{sec.results}, 
it is worth looking ahead to anticipate possible future developments.

One immediate development would be to explore excited polaron states beyond the ground state.
This will require us to solve Eq.~(\ref{eq:basis.4}) for higher-lying electronic eigenstates
instead of retaining only the ground state. The study of electronic excitations at fixed
lattice distortion could be useful to understand the response of polarons to ultrafast
optical excitations for example.

Another important development would be to go beyond the adiabatic and classical approximations.
Indeed, the main limitation of the present approach is that the starting point of the formalism
is the DFT energy functional in Eq.~(\ref{eq:DFT.1}). In this functional the electronic structure 
is described as a parametric function of classical ionic coordinates, therefore both DFT
calculations of polarons and our formalism are both similar in spirit to the Landau-Pekar polaron model.

Ideally we would want to study this problem using a fully-fledged field-theoretic formulation,
as provided for example by the self-consistent Hedin-Baym equations for the coupled 
electron-phonon system.\cite{Giustino2017} While it may be possible to proceed along this direction,
we speculate that it may be easier to start from the present formulation, and upgrade the
theory by reinstating from the outset non-adiabatic effects and quantum nuclear fluctuations.
For example, we could restart from Eq.~(\ref{eq:DFT.7}), introduce the quantum kinetic energy
of the nuclei, and write the problem in terms of the correlated electron-ion wavefunction
$\Psi(\br,\{\Delta \tau_{\k \a p}\})$:
  \begin{eqnarray}\label{eq.nonad}
   E_{\rm p}^\prime[\Psi] &=& 
  \int \! d\br\,d\{\Delta \tau_{\k \a p}\}\, \Psi^*(\br,\{\Delta \tau_{\k \a p}\}) \times \nonumber \\
   \Bigg[&-&\frac{1}{2}\sum_{\k\a p} \frac{\hbar^2}{2M_\k}\frac{\D^2}{\partial \Delta \tau_{\k \a p}^2} \nonumber \\[2pt]
  & +&
  \frac{1}{2}\sum_{\substack{\k\a p\\\k'\a' p'}} C^0_{\k \a p, \k'\a' p'} 
  \Delta\tau_{\k \a p} \Delta\tau_{\k' \a' p'} \nonumber \\[-2pt] 
  &+& \hH_{\rm KS}^0 + 
  \sum_{\k \a p} \frac{\D V_{\rm KS}^0 }{\D \tau_{\k \a p}} \Delta \tau_{\k \a p} \Bigg]
  \Psi(\br,\{\Delta \tau_{\k \a p}\}).\,\,
  \end{eqnarray}
The advantage of this formulation is that one could focus on a single electron interacting with a
phonon bath, because the electron-electron interaction is already captured by the DFT KS Hamiltonian.

Equation~(\ref{eq.nonad}) can be reformulated in terms of phonon ladder operators
and electron-phonon matrix elements, following steps similar to Sec.~\ref{sec.pol-eqs-h}.
Using the same notation as in Ref.~\onlinecite{Giustino2017}, the Hamiltonian inside the
square brackets becomes (apart from a constant provided by the zero-point energy):
  \begin{eqnarray}\label{eq.nonad2}
  \hat{H}_{\rm p}^\prime &=& \sum_{n\bk}\ve_{n\bk} |n\bk\>\<n\bk|
  + \sum_{\bq\nu} \hbar \w_{\bq\nu} \,\ha^{\dagger}_{\bq\nu}\ha_{\bq\nu} 
   \nonumber   \\[2pt]
  &+&N_p^{-\frac{1}{2}}  \!\!\!\!\! 
  \sum_{mn\nu,\bk,\bq} g_{mn\nu}(\bk,\bq) 
  (\ha_{\bq\nu}+\ha^{\dagger}_{-\bq\nu})  |m \bk+\bq\>\<n \bk|, \nonumber \\
  \end{eqnarray}
where the summations over bands are restricted to conduction or to valence states for
electron or hole polarons, respectively.
In this equation we do not employ the usual electron field operator
because we have only one electron, therefore second quantization only applies to phonons.

Equation~(\ref{eq.nonad2}) can be considered as the {\it ab initio} counterpart of the 
Fr\"ohlich electron-phonon Hamiltonian.\citep{Frohlich1954} In fact the standard Fr\"ohlich
Hamiltonian is recovered by retaining only one parabolic band and considering only one LO
phonon branch. This equation suggests a possible route to link the present approach
 with many-body calculations of model polaron Hamiltonians: (i) for a given system we could 
identify the most important electronic bands, phonons, and electron-phonon
couplings using our spectral decomposition into $A_{n\bk}$ and $B_{\bq\nu}$;
(ii) we could then simplify Eq.~(\ref{eq.nonad2}) to retain only the most important contributions;
(iii) at this point we could employ advanced many-body techniques for polaron
Hamiltonians, such as for example diagrammatic Monte Carlo (DMC) approaches.\citep{Prokofev1998}
In this way one could envision complete first-principles calculations of polarons,
where the atomistic details and predictive power of DFT approaches are combined with 
the wealth of many-body physics of DMC or other field-theoretic techniques.

\section{Summary}\label{sec.conclusions}

In this work we developed a first-principles methodology that enables calculations
of polaron energies and wavefunctions without using supercells. Our method
employs electronic band structures, phonon dispersion relations, and electron-phonon 
matrix elements calculated in the crystal unit cell using density-functional theory 
and density-functional perturbation theory. In our theory we formulate the polaron
problem as a variational minimization of a DFT functional including a self-interaction
correction for the polaron wavefunction. This strategy leads to a non-linear system
of two coupled equations for the electron or hole wavefunction and the associated atomic 
displacements. We showed that this approach has a mathematical structure similar to
the classic Landau-Pekar polaron problem, but in our case the coupling to all phonons,
both acoustic and optical, and both short- and long-range, is taken into account.

We applied this method to three test cases, namely the large electron polaron in
a halide salt, LiF, the small hole polaron in the same material, and the small
electron polaron in a layered metal oxide, Li$_2$O$_2$. In the case of the large
polaron we validated our calculation using the continuous Landau-Pekar model; in
the case of the small polaron we compared our results with explicit supercell
calculations. We observed that our technique describes correctly and accurately
both large and small polarons, therefore this method carries general validity 
across the length scales.

We introduced a spectral analysis of the polaron wavefunction and atomic displacements
in order to quantify which electron bands, phonon modes, and electron-phonon couplings
play the most important role in the formation of the polaron. This analysis allowed
us to identify Fr\"ohlich-type electron polarons in LiF, and Holstein-type polarons
in LiF (holes) and Li$_2$O$_2$ (electrons). We anticipated that this type of analysis will
be useful to devise model polaron Hamiltonian starting from
realistic materials parameters computed from first principles.
 
We hope that the present work will serve as the basis for future {\it ab initio}
calculations of polarons in real materials, and it will help combining together
the strengths of DFT-type calculations with field-theoretic polaron techniques
developed for model Hamiltonians.

\acknowledgments
This work was supported by a postgraduate scholarship from the Macau SAR Government (W.H.S.),
by the Leverhulme Trust (Grant RL-2012-001), the UK Engineering and Physical Sciences Research Council 
(grants No.  EP/L015722/1 and EP/M020517/1), the Graphene Flagship (Horizon 2020 Grant No. 785219 - 
GrapheneCore2), the University of Oxford Advanced Research Computing (ARC) facility 
(http://dx.doi.org/810.5281/zenodo.22558), the PRACE-17 resources MareNostrum at BSC-CNS.

\bibliography{cite.bib}

\newpage

\end{document}